\documentclass[fleqn,usenatbib]{mnras}

\usepackage{newtxtext,newtxmath}

\usepackage[T1]{fontenc}

\DeclareRobustCommand{\VAN}[3]{#2}
\let\VANthebibliography\thebibliography
\def\thebibliography{\DeclareRobustCommand{\VAN}[3]{##3}\VANthebibliography}

\usepackage{graphicx}	
\usepackage{amsmath}	
\usepackage{amssymb}	
\usepackage{xspace}
\usepackage{longtable}
\usepackage{subfig}
\usepackage{booktabs}
\usepackage{lscape}
\usepackage{siunitx}
\usepackage[symbol]{footmisc}

\newcommand{\Kepler}{\ensuremath{\emph{Kepler}}\xspace}

\newcommand{\astep}{\ensuremath{\emph{ASTEP}}\xspace}
\newcommand{\tess}{\ensuremath{\emph{TESS}}\xspace}
\newcommand{\jwst}{\ensuremath{\emph{JWST}}\xspace}
\newcommand{\ariel}{\ensuremath{\emph{Ariel}}\xspace}

\title[Three Neptunian planets orbiting HD~28109]{HD~28109 hosts a trio of transiting Neptunian planets including a near-resonant pair, confirmed by ASTEP from Antarctica}

\author[G. Dransfield et al.]{Georgina Dransfield,$^{1}$\thanks{E-mail: gxg831@bham.ac.uk}
Amaury H.M.J. Triaud,$^{1}$ 
Tristan Guillot,$^{2}$ 
Djamel Mekarnia,$^{2}$ 
David Nesvorn\'y,$^{3}$ 
\newauthor 
Nicolas Crouzet,$^{4,5}$ 
Lyu Abe,$^{2}$ 
Karim Agabi,$^{2}$ 
Marco Buttu,$^{6}$ 
Juan Cabrera,$^{7}$ 
Davide Gandolfi,$^{8}$ 
\newauthor 
Maximilian N. G{\"u}nther$^{4}$\thanks{ESA Research Fellow}, 
Florian Rodler,$^{9}$ 
Fran{\c c}ois-Xavier Schmider,$^{2}$ 
Philippe Stee,$^{2}$ 
Olga Suarez,$^{2}$ 
\newauthor 
Karen A. Collins,$^{10}$ 
Martín Dévora-Pajares,$^{11}$ 
Steve B. Howell,$^{12}$ 
Elisabeth C. Matthews,$^{13}$ 
\newauthor
Matthew R. Standing,$^{1}$ 
Keivan G.\ Stassun$^{14}$, 
Chris Stockdale,$^{15}$ 
Samuel N. Quinn,$^{10}$ 
Carl Ziegler,$^{16}$ 
\newauthor
Ian J.\ M.\ Crossfield,$^{17}$ 
Jack J. Lissauer,$^{11}$ 
Andrew W. Mann,$^{18}$ 
Rachel Matson,$^{19}$ 
Joshua Schlieder,$^{20}$ 
\newauthor
George Zhou$^{21}$ 
\\
\\
$^{1}$School of Physics \& Astronomy, University of Birmingham, Edgbaston, Birmingham B15 2TT, United Kingdom\\
$^{2}$Universit\'e C\^ote d'Azur, Observatoire de la C\^ote d'Azur, CNRS, Laboratoire Lagrange, Bd de l'Observatoire, CS 34229, 06304 Nice cedex 4, France\\
$^{3}$Southwest Research Institute, Department of Space Studies 1050 Walnut St., Suite 400, Boulder, Colorado 80302\\
$^{4}$European Space Agency (ESA), European Space Research and Technology Centre (ESTEC), Keplerlaan 1, 2201 AZ Noordwijk, The Netherlands\\
$^{5}$Leiden Observatory, Leiden University, Postbus 9513, 2300 RA, Leiden, The Netherlands\\
$^{6}$INAF OAS, Osservatorio di Astrofisica e Scienza dello Spazio di Bologna via Piero Gobetti 93/3, Bologna, Italy\\
$^{7}$Deutsches Zentrum f{\"u}r Luft- und Raumfahrt, Rutherfordstr. 2, D-12489 Berlin, Germany\\
$^{8}$Dipartimento di Fisica, Universit\`{a} Degli Studi di Torino Via Pietro Giuria, 1, 10125, Torino, Italia \\
$^{9}$European Southern Observatory, Alonso de Cordova 3107, Vitacura, Santiago, Chile \\
$^{10}$Center for Astrophysics \textbar \ Harvard \& Smithsonian, 60 Garden St, Cambridge, MA 02138, USA\\
$^{11}$Dpto. Física Teórica y del Cosmos, Universidad de Granada, 18071, Granada, Spain\\
$^{12}$NASA Ames Research Center, Moffett Field, CA 94035, USA \\
$^{13}$Geneva Observatory, University of Geneva, Chemin Pegasi 51, 1290 Versoix, Switzerland \\
$^{14}$Department of Physics \& Astronomy, Vanderbilt University, 6301 Stevenson Center Ln., Nashville, TN 37235, USA \\
$^{15}$Hazelwood Observatory\\
$^{16}$Department of Physics, Engineering and Astronomy, Stephen F. Austin State University, 1936 North St, Nacogdoches, TX 75962, USA \\
$^{17}$Department of Physics and Astronomy, University of
  Kansas, Lawrence, KS, USA \\
$^{18}$Department of Physics and Astronomy, The University of North Carolina at Chapel Hill, Chapel Hill, NC 27599-3255, USA \\
$^{19}$The United States Naval Observatory, 3450 Massachusetts Ave, NW, Washington, DC 20392-5420 \\
$^{20}$NASA Goddard Space Flight Center, 8800 Greenbelt Rd, Greenbelt, MD 20771, United States \\
$^{21}$Centre for Astrophysics, University of Southern Queensland, West Street, Toowoomba, QLD 4350, Australia
}

\date{Accepted 2022 May 5. Received 2022 April 11; in original form 2021 November 12}

\pubyear{2022}

\begin{document}
\label{firstpage}
\pagerange{\pageref{firstpage}--\pageref{lastpage}}
\maketitle

\begin{abstract}
We report on the discovery and characterisation of three planets orbiting the F8 star HD~28109, which sits comfortably in \tess's continuous viewing zone. The two outer planets have periods of $\rm 56.0067 \pm 0.0003~days$ and $\rm 84.2597_{-0.0008}^{+0.0010}~days$, which implies a period ratio very close to that of the first-order 3:2 mean motion resonance, exciting transit timing variations (TTVs) of up to $\rm 60\,mins$. These two planets were first identified by \tess, and we identified a third planet in the \textcolor{black}{\tess photometry} with a period of $\rm 22.8911 \pm 0.0004~days$. We confirm the planetary nature of all three planetary candidates using ground-based photometry from {\it Hazelwood}, \astep and \textit{LCO}, including a full detection of the $\rm \sim9\,h$ transit of HD~28109 c from Antarctica. The radii of the three planets are \textcolor{black}{$\rm R_b=2.199_{-0.10}^{+0.098} ~R_{\oplus}$, $\rm R_c=4.23\pm0.11~ R_{\oplus}$ and $\rm R_d=3.25\pm0.11 ~R_{\oplus}$}; we characterise their masses using TTVs and precise radial velocities from ESPRESSO and HARPS, and find them to be $\rm M_b=18.5_{-7.6}^{+9.1}~M_{\oplus}$, $\rm M_c=7.9_{-3.0}^{+4.2}~M_{\oplus}$ and $\rm M_d=5.7_{-2.1}^{+2.7}~M_{\oplus}$, making planet b a dense, massive planet while c and d are both under-dense. We also demonstrate that the two outer planets are ripe for atmospheric characterisation using transmission spectroscopy, especially given their position in the CVZ of {\it JWST}. The data obtained to date are consistent with resonant (librating) and non-resonant (circulating) solutions; additional observations will show whether the pair is actually locked in resonance or just near-resonant.
\end{abstract}




\section{Introduction}

The discovery of over $4\,500$ extrasolar planets has in no way diminished our curiosity regarding our place in the universe; if anything, we have more questions than ever. 

Following early results by Doppler surveys \citep[e.g.][]{2011arXiv1109.2497M, 2010ApJ...721.1467H}, the \Kepler survey \citep{2010Sci...327..977B} revealed an abundant population of planets comparable in size to Neptune \citep{2013ApJ...766...81F}, the majority occupying  periods shorter than a few hundred days 
\citep{2010Sci...330..653H}. The formation of these Neptune-size worlds remains debated; one possible pathway is that `pebbles' (large dust particles) separate from the gas in the planet-forming disc and drift inward \citep{2017AREPS..45..359J}. Another possibility is that the pebbles are accreted into cores before migrating inward \citep{2016MNRAS.457.2480C}; thus both models predict that most of the mass growth takes place in situ \citep[e.g.][]{2019AAA...627A..83L, 2021JGRE..12606639B}.

Both the `drift' and `migration' formation pathways predict the formation of resonant chains, where adjacent planets find themselves in a mean-motion resonance with their neighbours \citep[][]{2021JGRE..12606639B}. However most planets are not found with commensurate orbital periods \citep{2014ApJ...790..146F}. Studies show that in $\sim$95\% of cases the resonance chain becomes unstable following the dispersal of the gas disc \citep[][]{2010ApJ...714..194M,2015ApJ...807...44P, 2021AAA...650A.152I}. Alternatively, the migratory pathway might be more chaotic \citep{2012MNRAS.427L..21R, 2015ApJ...811...41L}. 

Statistical studies of the large samples of known planets have shown that multiplicity is common, i.e. many planets are in multi-planet systems \citep{2014ApJ...790..146F}. Those planets that find themselves close to a mean-motion resonance with their neighbours allow for masses to be constrained without spectroscopic follow-up, by monitoring transit timing variations (TTVs) in the system \citep[e.g.][]{2005MNRAS.359..567A, 2005Sci...307.1288H, 2010Sci...330...51H, 2012ApJ...761..122L}. 

Neptune-sized planets orbiting bright stars provide excellent dynamical laboratories to further investigate these exciting systems. \tess, the Transiting Exoplanet Survey Satellite \citep{2015JATIS...1a4003R} launched in 2018 to search for exoplanet candidates; one of its primary mission aims is to deliver to the community 50 planets  smaller than $4~\rm R_{\oplus}$, with masses measured by high precision radial velocity (RV). Not all systems are suitable for RV mass measurements, but it is predicted that \tess will also observe $\sim30$ systems with TTVs, with approximately a third of those being suitable for dynamical mass measurements \citep{2019AJ....158..146H}. 
Another of \tess's goals is the identification of planets orbiting bright stars, to permit detailed investigations into the dynamics and chemical composition of exoplanetary atmospheres. The discovery and analysis of HD~28109 (TOI-282) fulfils two of those primary objectives.

Stars within \tess's northern and southern continuous viewing zones (CVZ) are continuously observed for a year, allowing for the discovery of long period planets ($>20$ days). These planets are likely to be favourable candidates for {\it JWST} as its CVZ falls within \tess's; however they are challenging to confirm with ground-based photometry, especially if TTVs introduce large timing uncertainties.

Unless, of course, one is observing from Antarctica. \astep \citep[{\bf A}ntarctic {\bf S}earch for {\bf T}ransiting {\bf E}xo{\bf P}lanets][]{2015AN....336..638G, 2016MNRAS.463...45M} is a $40~\rm cm$ telescope located on the Antarctic Plateau at a latitude of $-75.1 \deg$ an elevation of $3,233~\rm m$. This unique location has a thin atmosphere, completely dry air, and virtually uninterrupted observing between late May and late July \citep{2010EAS....40..367C, 2010AAA...511A..36C}. The visibility covers \tess's CVZ and the area where Low-Earth Orbit telescopes have trouble reaching \citep[e.g. CHEOPS][]{2013EPJWC..4703005B}. For this reason, \astep is the observatory (ground or space) best suited for the photometric confirmation and follow-up of long-period transiting systems. 

In this work we present the discovery, validation and characterisation of three sub-Neptunes orbiting HD~28109. We begin by characterising the host in Section \ref{sec:starchar}, followed by Sections \ref{sec:cands} and \ref{sec:validation} where we describe the identification and validation of planetary candidates in the system. In Section \ref{sec:analysis} we describe our global analysis of all available photometric and radial velocity data to characterise the planets. We then place the three planets in the context of the field, including potential for atmospheric characterisation of the system and further high precision radial velocity follow-up in Section \ref{sec: architecture}. Finally, we summarise our findings and conclude in Section \ref{sec:282conc}.

\section{Stellar Characterisation}
\label{sec:starchar}

HD~28109 (TOI-282) is a bright (V=9.42; J = 8.476) main-sequence star, of spectral type F8/G0V. Its right ascension and declination are 04:20:57.19 -68:06:09.68, and it has a parallax of \textcolor{black}{$7.13~\rm mas$ \citep{2021AAA...649A...1G, 2016AAA...595A...1G}}, placing it at a distance of \textcolor{black}{$140~\rm pc$ \citep{2021AJ....161..147B}}.

All planetary measurements are derived from our knowledge of the star; we therefore begin by characterising the radius, mass, effective temperature and spectral type of the star. 

\subsection{Reconnaissance Spectroscopy}


We observed TOI-282 on three nights spanning 2019 February 14 to 2021 January 07 with the CHIRON spectrograph on the 1.5 m SMARTS telescope. CHIRON is a high resolution echelle spectrograph fed by an image slicer and a fiber bundle, located at Cerro Tololo Inter-American Observatory (CTIO), Chile. The spectra have a resolution of $\rm R\,=\,80000$ with a wavelength coverage of $4500-8900\,\si{\angstrom}$ \citep{2013PASP..125.1336T}. The wavelength solution is provided by bracketing Thorium-Argon cathode-ray lamp observations, and the spectra are extracted and wavelength calibrated with the official CHIRON pipeline \citep{2021AJ....162..176P}. We extracted the RVs by fitting the spectral line profiles, which were measured via least-squares deconvolution of the observed spectra using synthetic templates \citep{1997MNRAS.291..658D}. \textcolor{black}{The three RVs exhibit a root mean square scatter of $58\,m/s$, which is not significant given the $44\,m/s$\ mean per-point uncertainty. With these data, we cannot place strong limits on the presence of additional companions, but the two-year time span of the observations does allow us to measure long-term trends that might be indicative of very massive outer companions. The best-fit linear trend ($0.076 \pm 0.091\,m/s/day$) is consistent with zero, which suggests it is unlikely that there is a stellar companion in the system, but the precise radial velocities described in Section \ref{sec:rv_intro} can provide tighter constraints on the masses of the transiting planets and limits on the presence of other bound companions.} 

We also use the CHIRON spectra to determine the effective temperature, surface gravity, and metallicity of the host star by matching against observed spectra that have previously been classified using SPC \citep{2012Natur.486..375B}. Interpolation to the final parameters is performed with a gradient-boosting regressor implemented in the \textsc{scikit-learn} python module. We measure $T_{\rm eff} = 6120 \pm 50$\ K, $\log{g_\star} = 4.13 \pm 0.10$, and $\left[{\rm Fe}/{\rm H}\right] = 0.0 \pm 0.10$. Following \citet{2005oasp.book.....G} and \citet{2018AJ....156...93Z}, we derive $v\sin{i_\star}$ by fitting broadening kernels to the instrumental, macroturbulent, and rotational line profiles, and estimate $v\sin{i_\star} = 7.7 \pm 0.5\,km\,s^{-1}$.

\subsection{Spectral Energy Distribution}
\label{SED}

\begin{figure}
    \centering
    \includegraphics[width =\columnwidth]{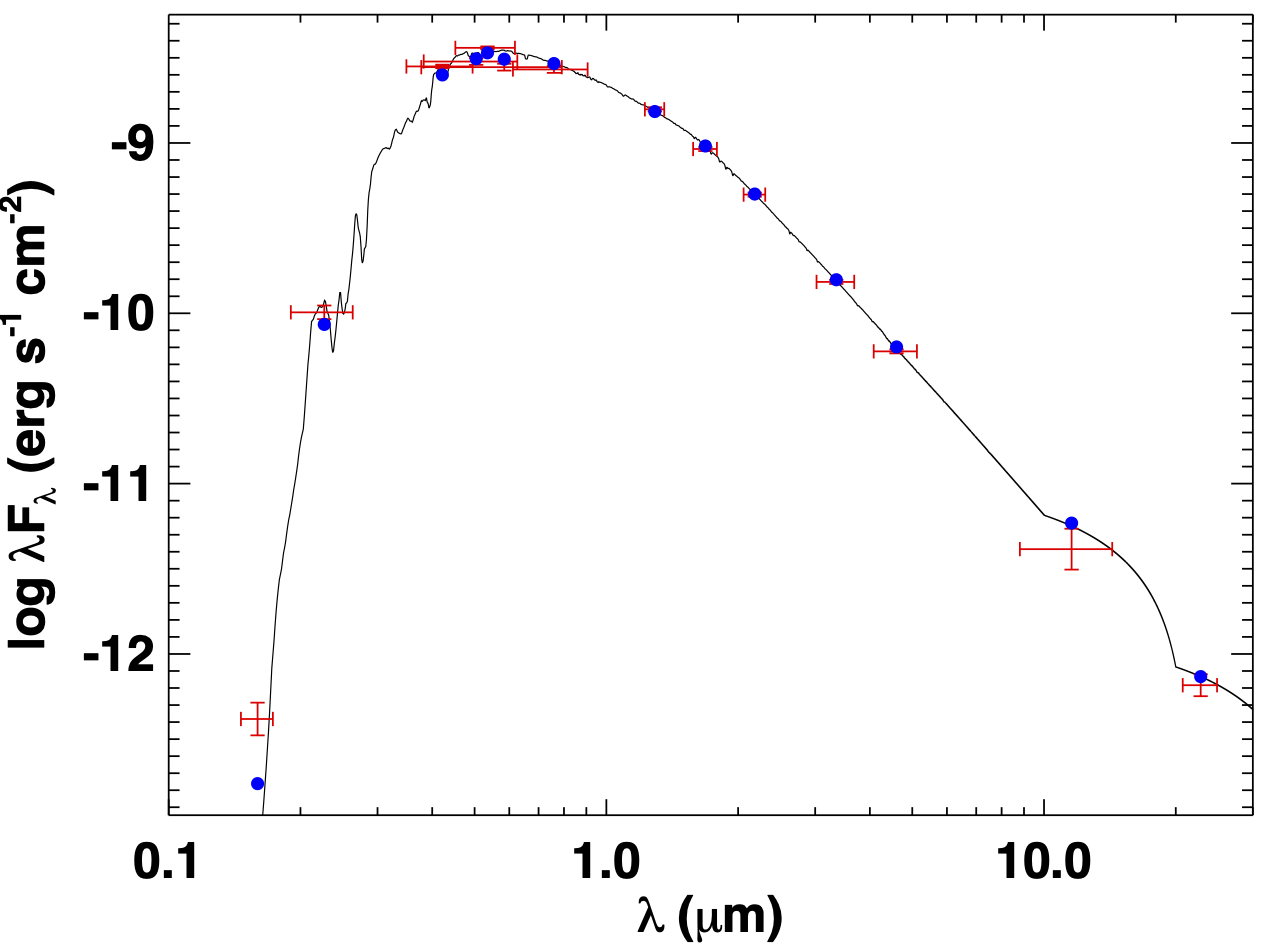}
    \caption{Spectral energy distribution (SED) of TOI-282. Red symbols represent the observed photometric measurements, where the horizontal bars represent the effective width of the passband. Blue symbols are the model fluxes from the best-fit Kurucz atmosphere model (black).}
    \label{fig:sed}
\end{figure}

As an independent check on the derived stellar parameters, and in order to determine an estimate for stellar age, we perform an analysis of the broadband spectral energy distribution (SED). Together with the {\it Gaia\/} EDR3 parallax, we determine an empirical measurement of the stellar radius following the procedures described in \citet{Stassun:2016,Stassun:2017,Stassun:2018}. We pulled the $B_T V_T$ magnitudes from {\it Tycho-2}, the $JHK_S$ magnitudes from {\it 2MASS}, the W1--W4 magnitudes from {\it WISE}, and the $G G_{\rm RP} G_{\rm BP}$ magnitudes from {\it Gaia}. In addition, we pulled the FUV and NUV fluxes from {\it GALEX\/} in order to assess the level of chromospheric activity, if any. Together, the available photometry spans the full stellar SED over the wavelength range 0.2--22~$\mu$m (see Figure~\ref{fig:sed}).  

We perform a fit using Kurucz stellar atmosphere models \citep[][]{1979ApJS...40....1K}, with the effective temperature ($T_{\rm eff}$) and metallicity ([Fe/H]) adopted from the spectroscopic analysis. The remaining free parameter is the extinction, $A_V$, which was limited to the maximum line-of-sight value from the dust maps of \citet{Schlegel:1998}. The resulting fit is very good (Figure~\ref{fig:sed}) with a best-fit $A_V = 0.01\pm0.01$ and a reduced $\chi^2$ of 2.8 (excluding the {\it GALEX\/} FUV flux, which is consistent with a modest level of chromospheric activity; see below). Integrating the (unreddened) model SED gives the bolometric flux at Earth of $F_{\rm bol} = 4.37 \pm 0.15 \times 10^{-9}$ erg~s$^{-1}$~cm$^{-2}$. Taking the $F_{\rm bol}$ and $T_{\rm eff}$ together with the {\it Gaia\/} EDR3 parallax, with no systematic offset applied \citep[see, e.g.,][]{StassunTorres:2021}, gives the stellar radius as $R_\star = 1.446 \pm 0.035$~R$_\odot$. Finally, estimating the stellar mass from the empirical relations of \citet{Torres:2010} and a 6\% error from the empirical relation itself gives $M_\star = 1.26 \pm 0.08$~M$_\odot$, consistent with the mass estimated empirically from the stellar radius together with the spectroscopic $\log g$ which gives $M_\star = 1.03 \pm 0.24$~M$_\odot$. 

We can also estimate the stellar age by taking advantage of the observed modest chromospheric activity in the UV together with empirical age-activity-rotation relations. For example, we can estimate the $\log R'_{\rm HK} = -4.56 \pm 0.05$ via the {\it GALEX\/} FUV excess and the empirical relations of \citet{Findeisen:2011}. 
That implies an age of $\tau = 0.9 \pm 0.3$~Gyr via the empirical activity-age relations of \citet{Mamajek:2008}. 
Finally, we can further corroborate the activity-based age estimate by also using empirical relations to predict the stellar rotation period from the activity. For example, the empirical relation between $R'_{\rm HK}$ and rotation period from \citet{Mamajek:2008} predicts a rotation period for this star of $7.2 \pm 1.1$~d, which is compatible with the (projected) rotation period $P_{\rm rot}/\sin i_\star = 9.51 \pm 0.62$~d inferred from the stellar radius above together with the spectroscopic $v\sin i_\star$. 

We present the adopted stellar parameters in Table \ref{tab:starpar}

\begin{table}
\centering
\begin{tabular}{@{}lp{25mm}p{30mm}@{}}
\toprule
{\bf Designations} & \multicolumn{2}{p{65mm}}{HD~28109, TIC 29781292, TOI-282, 2MASS J04205712-6806095, UCAC4 110-003794, WISE J042057.17-680609.5, Gaia DR2 4668163021600295552, HIP 20295, TYC 9154-01248-1} \\ \midrule
{\bf Parameter} & {\bf Value}              & {\bf Source} \\ \midrule
T mag           & 8.9387$\pm$0.006        & \cite{2019AJ....158..138S} \\
B mag           & 9.91$\pm$0.03       & \cite{2000AAA...355L..27H} \\
V mag           & 9.38$\pm$0.02        & \cite{2000AAA...355L..27H} \\
\textcolor{black}{G mag}           & \textcolor{black}{9.3063$\pm$0.0028}        & \textcolor{black}{\cite{2021AAA...649A...1G}} \\
J mag           & 8.476$\pm$0.020          & \cite{2003yCat.2246....0C} \\
H mag           & 8.256$\pm$0.024          & \cite{2003yCat.2246....0C} \\
K mag           & 8.175$\pm$0.023          & \cite{2003yCat.2246....0C} \\
W1 mag           & 8.136$\pm$0.023          & \cite{2010AJ....140.1868W} \\
W2 mag           & 8.175$\pm$0.020          & \cite{2010AJ....140.1868W} \\
W3 mag           & 8.155$\pm$0.020            & \cite{2010AJ....140.1868W} \\
W4 mag           & 7.997$\pm$0.161        & \cite{2010AJ....140.1868W} \\
\textcolor{black}{Distance}           & \textcolor{black}{140.087$\pm$0.194}        & \textcolor{black}{\cite{2021AJ....161..147B}} \\
SpT             & F8/G0V                   & \cite{1975mcts.book.....H} \\
$R_{\star}$     & $1.446\pm0.035R_{\odot}$ & This work                  \\
$M_{\star}$     & $1.26\pm0.08M_{\odot}$   & This work                  \\
Age             & 1.1$\pm$0.1 Gyr             & This work                  \\
${\rm T_{eff}}$ & 6120$\pm$50 K             & This work                  \\
$\log g_\star$           & 4.13$\pm$0.10            & This work                  \\
$\rm [Fe/H]$          & 0.0$\pm$0.1              & This work                  \\
$v \sin i_\star$         & 7.7$\pm$0.5 km/s             & This work                  \\ \bottomrule
\end{tabular}
\caption{Stellar parameters adopted for this work.}
\label{tab:starpar}
\end{table}

\section{Identification of Planetary Candidates}
\label{sec:cands}


HD~28109 (TOI-282) is in the southern continuous viewing zone (CVZ) for \tess which means it is observed in all southern sectors, the only exception being sector 32 where the target was off the edge of the CCD by 3 pixels. At the time of writing, the star has been observed in 25 sectors at two-minute cadence.

In the following sections we describe the identification of candidates by \tess, as well as our own search for further candidates in the data.

\subsection{Note on Nomenclature}

Throughout Section \ref{sec:cands} we refer to the host (HD~28109) using its \tess alias: TOI-282, where TOI is \tess Object of Interest. When referring to candidate planets, we also use the \tess nomenclature and add numeric suffixes in the order that candidates were identified. Once the candidates are confirmed we change their names. Thus, TOI-282.01 becomes HD~28109~c; TOI-282.03 becomes HD~28109~d; and TOI-282.04 is HD~28109~b, with letters indicating the order from closest to farthest from the host star.

 \subsection{\tess data}
 \label{sec:tess}
 
 \begin{figure*}
    \centering
    \includegraphics[width =\textwidth]{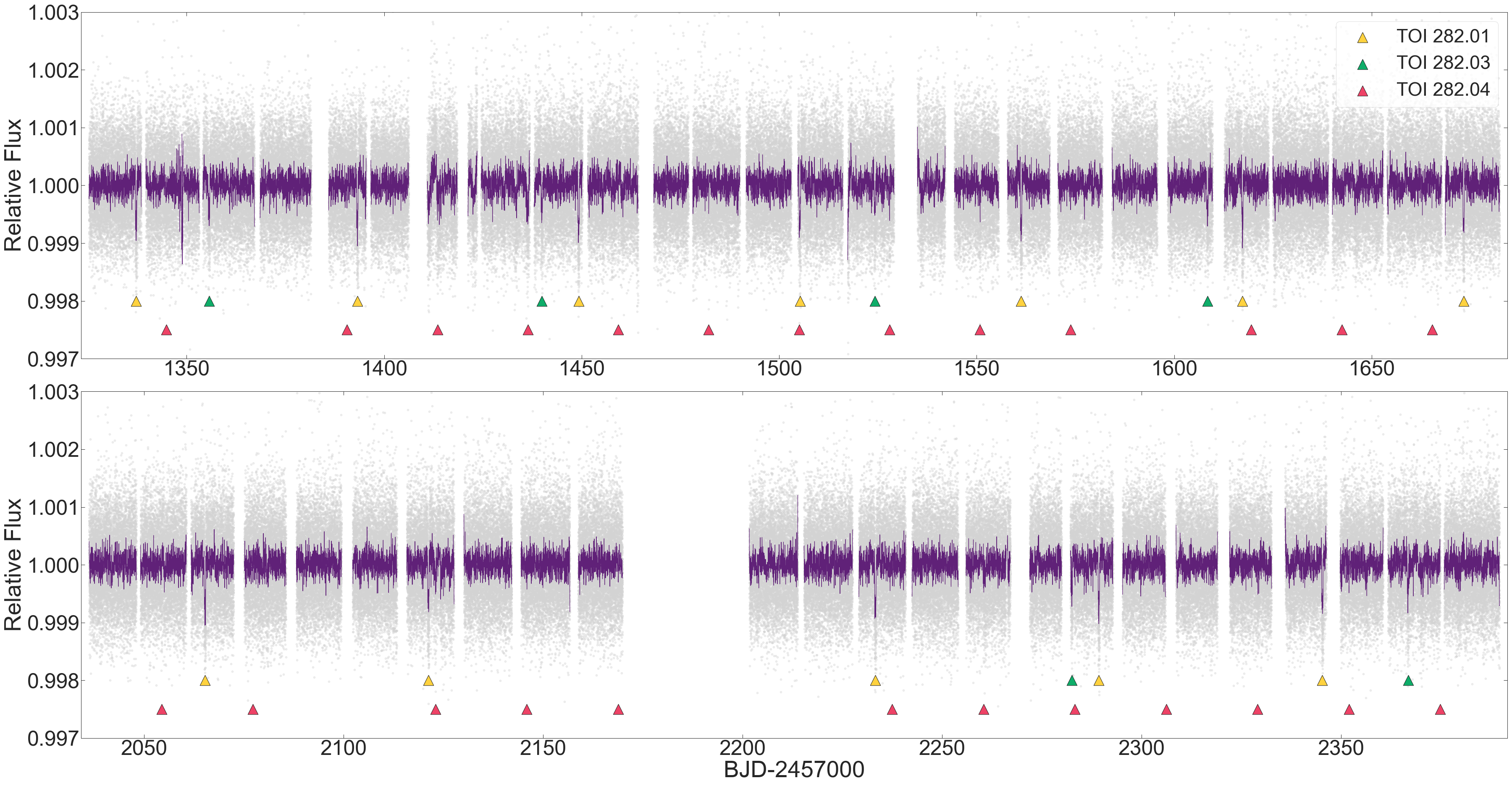}
    \caption{PDCSAP lightcurves of TOI-282, as produced by the SPOC pipeline. The top panel shows the photometry for sectors 1-13, while the bottom panel contained sectors 27-37, with a gap where sector 32 was missing. Binned data with a cadence of $30~\rm min$ are shown in \textcolor{black}{purple}. }
    \label{fig:tess_lc}
\end{figure*}
 
All 2 minute cadence \tess data is reduced by the SPOC (Science Processing Operations Center) pipeline \citep{2016SPIE.9913E..3EJ} and the photometry data products are delivered in the form of simple aperture photometry (SAP) or Presearch Data Conditioning Simple Aperture Photometry (PDCSAP), where the latter has been corrected for instrument systematics. The lightcurves are then searched for transit-like signals; candidates that have a signal-to-noise ratio $\geq7$ are reported as threshold crossing events (TCEs).

The first TCE on TOI-282 was detected in Sector 1, and a second was detected in Sector 2. The multi-sector data validation report for Sectors 1-2 showed this as a planetary candidate with a period of $\sim$18 days. It wasn't until Sectors 1-6 were available that the individual candidates could be disentangled. The 1--6 multi-sector validation report reported three planet candidates on TOI-282: candidate .01 with a period of $\approx$56 days, .02 with a period of $\approx$31 days, and finally .03 with a period of $\approx$84 days. Candidate .02 was later retired as a false alarm\footnote{The 31-day transit event was caused by a combination of noise and confusion with the 84-day transit events in early TESS data.}.

The PDCSAP lightcurves for Sectors 1-13 and Sectors 27-39 can be found in Figure \ref{fig:tess_lc}, where transits of TOI-282.01 are highlighted in yellow, while the transits of TOI-282.03 are in green.

\subsection{Search For Additional Transiting Candidates}
\label{sec:cand_search}

We made use of the custom \textsc{Sherlock}\footnote{\textsc{Sherlock} is publicly available at \url{https://github.com/franpoz/SHERLOCK}.} 
pipeline presented in \cite{2020AAA...641A..23P} and \cite{2020AAA...642A..49D} to perform a search for additional transiting candidates in the \tess data. \textsc{Sherlock} uses the \textsc{Lightkurve} package \citep{2018ascl.soft12013L} to download \tess PDCSAP data from NASA Mikulski Archive for Space Telescope (MAST); any outliers, defined as any points over 3$\sigma$ above the running mean, are then removed. The pipeline then uses \textsc{Wotan} \citep{2019AJ....158..143H} to detrend the data using the bi-weight method testing several window lengths. In this case, we used six window sizes between 0.64 and 3.55 days in order to optimise signal detection efficiency (SDE). To aid in the detection of low SNR signals, \textsc{Sherlock} also optionally applies a a Savitzky–Golay filter \citep{1964AnaCh..36.1627S} to smooth the data and increase precision. 

The search for periodic signals is carried out using the \textsc{Transit Least Square (TLS)} package \citep{2019ascl.soft10007H} as it is optimised to search for periodic signals with transit-like shapes. We searched a wide parameter space, with periods ranging from 5 to 100 days setting the minimum signal-to-noise ratio (SNR) to 5. This is in contrast to the \textsc{SPOC} pipeline, which sets the minimum SNR to 7.

\begin{figure}
    \centering
    \includegraphics[width = \columnwidth]{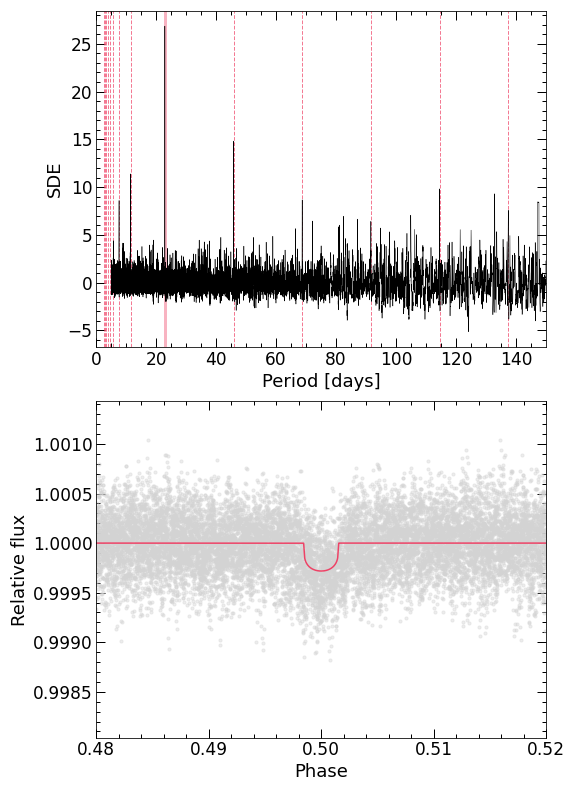}
    \caption{Results of our search for additional candidates in the data, using {\sc TransitLeastSquares} as implemented by {\sc Sherlock}. \textbf{Upper panel:} Lomb-Scargle periodogram showing the detected period and its harmonics. \textbf{Lower panel:} \tess lightcurve phase-folded on this period, with a transit model overplotted. }
    \label{fig:planet_d}
\end{figure}

We recover both candidates .01 and .03 in the first instance and find that the SDE is consistent across all window sizes, although the SNR and transit depths are at their highest using a window size of 2.8259 days. We also find a third significant signal at a period of 22.89$\pm$0.01 days. This signal has an SNR of 23.8 and SDE of 13.1 using the same window-size as for the first two candidates; its depth of just 0.14ppt (parts per thousand) would make it extremely challenging to detect from the ground. Nevertheless, if this is indeed a third planet in the system, its size would be classed as a mini-Neptune just beyond the so-called radius valley \citep[][]{2017AJ....154..109F, 2018MNRAS.479.4786V}, and its period would place it within $\sim5\%$ of a 5:2 third order resonance with candidate .01. The \tess lightcurve folded on this signal is presented in Figure \ref{fig:planet_d}, along with the Lomb-Scargle periodogram. 

\textcolor{black}{While this signal does not match any of the TOIs on ExoFOP, further scrutiny reveals that this might have been the period originally associated with TOI-282.02. \citet{2020AJ....160..107D} include the candidates of TOI-282 in their sample, and cite the orbital period for the innermost candidate as $\sim22.89~\rm days$. Additionally, when we scrutinise some of the first multi-sector data validation reports produced by the \textsc{SPOC} pipeline, we find that Candidate 2 has a period of $\sim22.89~\rm days$\footnote{All data validation reports for TOI-282 are publicly available at \url{https://exofop.ipac.caltech.edu/tess/target.php?id=29781292}. }}

We now adopt this re-identified candidate as TOI-282.04 and indicate in \textcolor{black}{dark pink} the detected transits on Figure \ref{fig:tess_lc}.

Two other periodic signals with low significance were also recovered which did not pass our vetting tests and are therefore attributed to instrument systematics.

\section{Vetting and Validation of Candidates}
\label{sec:validation}

In this section we describe first the ground-based follow-up observations we conducted, staring with high-resolution imaging of the host to search for blended companions. We then describe the photometric observations conducted from several sites between January 2019 and August 2021. In the first instance, the purpose of ground-based follow-up is to validate the planetary nature of the candidates; in Section \ref{sec:falsepos} we describe how we used the available data on this system to rule out all feasible false positive scenarios.

We note that while the validation tests described here allowed us to confirm the planetary nature of TOI-282.01, .03 and .04 initially, the analysis we describe in Section \ref{sec:dynamic} reveals anti-correlated transit timing variations between planets .01 and .03. Therefore, many of the tests here described are not necessary to validate these two objects as planets since TTVs are not known to have astrophysical false-positives. We nevertheless kept the validation steps here for completeness.

All follow-up observations are summarised in Table \ref{tab:followup}.

\begin{table*}
\centering
\caption{Summary of ground-based follow-up observations carried out for the validation of TOI-282.01, .03 and .04.}
\begin{tabular}{@{}ccccc@{}}
\multicolumn{5}{c}{\textbf{Follow-up Observations}}                                              \\ \midrule \midrule
\multicolumn{5}{c}{\textbf{High Resolution Imaging}}                                             \\ 
\textbf{Observatory} & \textbf{Filter} & \textbf{Date}     & \textbf{Sensitivity Limit} & \textbf{Result}\\ \midrule
{\it VLT}      & {\it $K_s$}     & 2019 January 25 & $\Delta m=5$ at $0.23-4.5"$  & No sources detected  \\
{\it SOAR}      & {\it $I_c$}     & 2019 February 02 &  $\Delta m=7.3$ at $1-3"$  & No sources detected   \\ 
{\it Gemini South}      & $562~{\rm nm}$     & 2020 December 23 & $\Delta m=5$ at $0.1-1.2"$  & No sources detected  \\ 
{\it Gemini South}      & $832~{\rm nm}$     & 2020 December 23 & $\Delta m=8$ at $0.1-1.2"$  & No sources detected   \\ \midrule
\multicolumn{5}{c}{\textbf{Photometric Follow-up}}                                               \\
\textbf{Observatory} & \textbf{Filter} & \textbf{Date}     & \textbf{Coverage} & \textbf{Result} \\ \midrule
\multicolumn{5}{c}{TOI-282.01 / HD~28109\,c}                                                                   \\ \midrule
{\it Hazelwood}      & {\it $I_c$}     & 2019 January 19 & Ingress $+ 50 \%$ & Field cleared of NEBs within $2.5"$\\
\astep               & {\it $R_c$}     & 2020 June 08     & Full              & Detection       \\
\astep               & {\it $R_c$}     & 2021 March 15   & Ingress $+ 50 \%$ & Detection       \\
\astep               & {\it $R_c$}     & 2021 May 10     & Ingress $+ 90 \%$ & Detection       \\
\astep               & {\it $R_c$}     & 2021 July 05    & Full              & Non-detection   \\
\astep               & {\it $R_c$}     & 2021 August 30  & Full              & Non-detection   \\ 
{\it LCO}               & {\it $z_s$}     & 2021 August 30  & Full              & Field cleared of NEBs  within $2.5"$   \\\midrule
\multicolumn{5}{c}{TOI-282.03  / HD~28109\,d}                                                                   \\ \midrule
\astep               & {\it $R_c$}     & 2021 March 08    & Ingress $+ 30 \%$ & Field cleared of NEBs  within $2.5"$ \\
\astep               & {\it $R_c$}     & 2021 August 23  & Ingress $+ 40 \%$ & Field cleared of NEBs  within $2.5"$   \\ \midrule
\multicolumn{5}{c}{TOI-282.04  / HD~28109\,b}                                                                   \\ \midrule
{\it Hazelwood}      & {\it $I_c$}     & 2019 January 19 & Ingress $+ 50 \%$ & Field cleared of NEBs  within $2.5"$\\ \midrule
\multicolumn{5}{c}{\textbf{Spectroscopic Observations}}                                               \\ 
\textbf{Instrument} & \textbf{Wavelength Range} & \textbf{Date Range}     & \textbf{Number of Spectra} & \textbf{Use}\\ \midrule
{\it CHIRON}      & $450-890~\rm nm$      &  2019 February 14 - 2021 January 07 & 3  & Stellar characterisation  \\
{\it ESPRESSO}      & $380-788~\rm nm$      &  UT 2019 October 15 - UT 2019 December 27  & 8  & Rule out stellar companion  \\
{\it HARPS}      & $383-693~{\rm nm}$     & UT 2019 May 24 - UT 2021 January 29  & 7  & Rule out stellar companion  \\

\end{tabular}
\label{tab:followup}
\end{table*}

\subsection{High Resolution Imaging}
\label{sec:speckle}

If an exoplanet star has a close companion (bound or line of sight), ``third-light” flux from the companion can lead to an underestimated planetary radius if not accounted for in the transit modelling \citep[]{2015ApJ...805...16C, 2017AJ....154...66F, 2018AJ....156...31M}. 

\subsubsection{Gemini South Telescope}

To search for close-in (bound) companions unresolved in other follow-up observations, we obtained high-resolution imaging observations from Gemini South’s Zorro speckle interferometric instrument\footnote {\url{https://www.gemini.edu/sciops/instruments/alopeke-zorro/}}.
 
TOI-282 was observed using Zorro on UT 2020 December 23. Zorro provides simultaneous high-resolution speckle imaging in two optical bands, 562/54 and 832/40 nm, with output data products including a reconstructed image, and robust limits on companion detections \citep[]{2011AJ....142...19H, 2021AJ....161..164H}. Figure \ref{fig:speckle} (upper panel) shows our 5$\sigma$ detection limit contrast curves and the corresponding reconstructed speckle image in 832 nm. We find that TOI-282 is a single star with no companions detected in our observations down to a contrast level of 5 to 8 magnitudes from the Gemini 8-m diffraction limit out to 1.2”. At the distance of TOI-282 (d=140 pc) these angular limits correspond to spatial limits of 2.8 to 168 AU.

\subsubsection{Southern Astrophysical Research Telescope}

We also searched for stellar companions to TOI-282 with speckle imaging on the 4.1-m Southern Astrophysical Research (SOAR) telescope \citep{2018PASP..130c5002T} on UT 2019 February 2, observing in Cousins I-band, a similar visible bandpass as \tess. This observation was sensitive to a 7.3-magnitude fainter star at an angular distance of 1 arcsec from the target. More details of the observation are available in \cite{2020AJ....159...19Z}. The 5$\sigma$ detection sensitivity and speckle auto-correlation functions from the observations are shown in Figure \ref{fig:speckle} (middle panel). No nearby stars were detected within 3\arcsec of TOI-282 in the SOAR observations.

\subsubsection{VLT NaCo}

\begin{figure}[h!]
    \centering
    \includegraphics[width=\columnwidth]{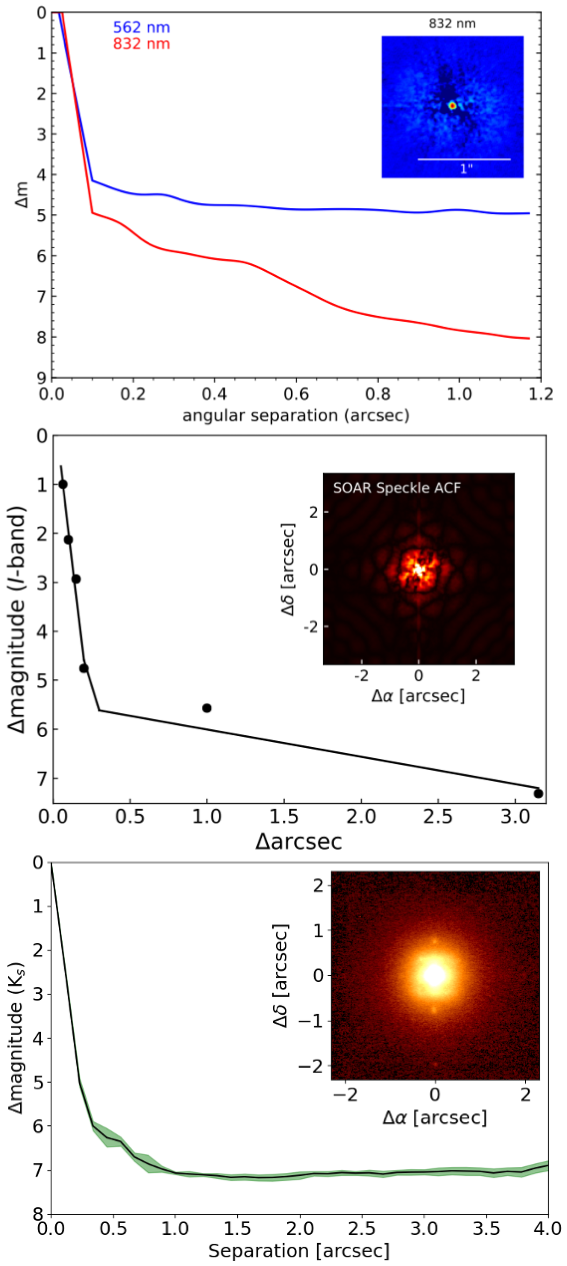}
    \caption{Contrast curves for HD~28109/TOI-282. The top panel shows the results from Gemini South, the middle panel shows the results from SOAR, and the bottom panel shows results from VLT/NaCo. All three panels are inset with final reconstructed images; none of the observations revealed bound companions within 5 magnitudes of the target. }
    \label{fig:speckle}
\end{figure}

We collected high-resolution AO images of TOI-282 with VLT/NaCo \citep{2003SPIE.4841..944L, 2003SPIE.4839..140R} on UT 2019 January 25, using the $K_s$ filter. We collected 9 frames, each with exposure time 20s, and dithered the telescope by 2'' in a grid-like pattern between each frame, and constructed a sky background frame by median combining the science images. We processed the data using a custom pipeline which corrects bad pixels, subtracts the sky frame and applies a flat-field correction, and then aligns the stellar position between each frame and co-adds the images. To test the sensitivity of these images, we injected faint fake PSFs into the data, and scaled them such that they could be redetected at $5\sigma$. Significances are averaged radially and presented in Figure \ref{fig:speckle} (lower panel); the data are sensitive to companions 5mag fainter than the star beyond 230mas, and to companions 7mag fainter than the star in the background limited regime. We searched for companions in the reduced images by eye. Detector persistence at the dither positions causes a faint point source 2'' directly to the south of the companion, but by inspecting the individual images we confirm that this is not a true visual companion. Apart from this persistence, no point sources are seen anywhere in the field of view, and we confirm that the star is single to the limit of our resolution.

\subsection{Photometric Follow-up}
\label{sec:followup}

Both candidates identified by \tess (TOI-282.01 and .03)  have long periods ($\rm \sim56\,d$ and $\rm \sim84\,d$), long transit durations ($\rm >8\,h$), and their transits are shallow. These constraints combined made TOI-282 a challenging system to follow-up from the ground, accounting for why the .01 only has seven photometric follow-up observations to date with three detections, while the .03 candidate has two observations with no transit detections. 

Below we describe first the observations carried out by {\it Hazelwood}, followed by the observations taken by \astep and \textit{LCO}. 

\subsubsection{Hazelwood}
\label{sec:hazel}

The {\it Hazelwood} Observatory is a backyard observatory with a $\rm 0.32\,m$ Planewave CDK telescope working at f/8, a SBIG STT3200 $\rm 2.2k \times 1.5k$ CCD, giving a $20\arcmin \times 13\arcmin$ field of view and $0.55\arcsec$ per pixel. The camera is equipped with $B, V, Rc, Ic, g', r', i'$ and $z'$  filters. Typical FWHM is 2.2" to 2.7". The Hazelwood Observatory, operated by Chris Stockdale in Victoria, Australia, observed an ingress of TOI-282.01 in Ic on UT 2019 January 21; \textcolor{black}{the transit was below the detection threshold, resulting in a flat lightcurve}, but a nearby eclipsing binary (NEB) check cleared the field within $2.5\arcmin$. 

Coincidentally, this observation spanned a rare double transit that included the innermost candidate (TOI-282.04) as well. The successful NEB check therefore also applies to this candidate, ruling out nearby stars as the source of the event.

\subsubsection{ASTEP}


\astep is a  custom 0.4\,m Newtonian telescope equipped with a 5-lens Wynne coma corrector and a 4k $\times$ 4k front-illuminated FLI Proline KAF-16801E CCD. The camera has an image scale of 0.''93 pixel$^{-1}$ resulting in a 1$^ {o} \times 1^ {o}$ corrected field of view. The focal instrument dichroic plate splits the beam into a blue wavelength channel for guiding, and a non-filtered red science channel roughly matching an $R_c$ transmission curve \citep{2013AAA...553A..49A, 2015AN....336..638G}. The telescope is automated or remotely operated when needed. Due to the extremely low data transmission rate at the Concordia Station, the data are processed on-site using an automated IDL-based aperture photometry pipeline \citep{ 2016MNRAS.463...45M}. The calibrated light curve is reported via email and the raw light curves of about 1,000 stars of the field are  transferred to Europe on a server in Roma, Italy and are then  available for deeper analysis. These data files contain each star’s flux computed through $10$ fixed circular apertures radii, so that optimal light curves can be extracted.

The \astep location at Dome C, on the Antarctic plateau,  means that it enjoys excellent photometric conditions and exceptionally long nights \citep{2018AAA...619A.116C}. \astep is therefore well placed to observe targets with long orbital periods and transit durations.


We observed a transit of the candidate TOI-282.01 with \astep on the night of UT 2020 June 08 following our interest to confirm planets with long transits that we were particularly well able to observe during the austral winter. TOI-282.01 was identified by searching the \textsc{TESS Transit Finder} \citep{2013ascl.soft06007J}. Our detection of the full transit, the first of its kind as part of the photometric follow-up effort of TFOP (\tess Follow-up Observing Program) Sub-Group 1, prompted us to follow the system and look more closely at the available \tess data. 

TOI-282.01 was observed on four further occasions by \astep (UT 2021 March 15, UT 2021 May 10, UT 2021 July 05, UT 2021 August 30) resulting in two partial transits and two non-detections due to poor weather. 

TOI-282.03 was observed on the UT 2021 March 08 and again on the UT 2021 August 23. Both observations resulted in non-detections of the shallow event, but the field was cleared of NEBs within 2.5".

\subsubsection{LCO}

We observed TOI-282 in Pan-STARRS $z$-short band on UTC 2021 August 30 from the Las Cumbres Observatory Global Telescope \citep[LCOGT;][]{Brown:2013} 1.0\,m network node at Siding Spring Observatory. The 1\,m telescopes are equipped with $4096\times4096$ SINISTRO cameras having an image scale of $0\farcs389$ per pixel, resulting in a $26\arcmin\times26\arcmin$ field of view. The images were calibrated by the standard LCOGT {\sc BANZAI} pipeline \citep{McCully:2018}, and photometric data were extracted using {\sc AstroImageJ} \citep{Collins:2017}. The target star was saturated to provide higher photometric precision for fainter \textcolor{black}{stars} nears within $2\farcs5$ of the target star. The observations were scheduled to include ingress coverage for a transit of TOI-282.03 with more than $\pm3\sigma$ ephemeris coverage according to a public ephemeris from the SPOC pipeline. However, according to the linear ephemeris extracted in Section \ref{sec:glob_phot} of this work, the observations covered from the time of nominal ingress to 90 minutes later. According to the TTV analysis of Section \ref{sec:dynamic}, the ingress would have happened 45 minutes earlier than the linear ephemeris from this work on this epoch, resulting in all in-transit coverage. Although these data do not rule out nearby eclipsing binaries due to the ingress timing, we do not find an obvious deep event in any star within $2\farcs5$ of the target star that could be the source of the TESS signal.

\subsection{Radial Velocity Follow-up}
\label{sec:rv_intro}

We acquired 8 high-resolution spectra (R$\approx$140\,000) with the Echelle SPectrograph for Rocky Exoplanets and Stable Spectroscopic Observations  \citep[ESPRESSO;][]{Pepe2021} on the 8.2\,m ESO Very Large Telescope (VLT; Paranal observatory, Chile). The observations were performed in service mode between UT 2019 October 15 and UT 2019 December 27 as part of our program 0104.C-0003 (PI: Rodler). The exposure time was set to 1115 s, leading to a mean signal-to-noise (S/N) ratio of $\sim$160 per pixel at 550\,nm. We reduced the ESPRESSO data and extracted the radial velocity measurements by employing ESO's ESPRESSO data reduction pipeline, version 2.2.1\footnote{\url{https://www.eso.org/sci/software/pipelines/espresso/espresso-pipe-recipes.html}}, \textcolor{black}{reaching a mean RV precision of $\rm 1.1\,m/s$}.

We also gathered 7 high-resolution spectra (R$\approx$115\,000) with the High Accuracy Radial Velocity Planet Searcher \citep[HARPS;][]{Mayor2003} spectrograph mounted at the 3.6\,m ESO telescope (La Silla observatory, Chile), as part of ESO programs 60.A-9700 and 60.A-9707. The spectra were collected between UT 2019 May 24 and UT 2021 January 29, setting the exposure time to 1800 s, which led to a mean S/N ratio of $\sim$100 per pixel at 550\,nm. We reduced the spectra using the dedicated HARPS data reduction software (DRS) and extracted the radial velocity (RV) measurements by cross-correlating the Echelle spectra with a G2 numerical mask \citep{Baranne1996,Pepe2002, Lovis2007}. \textcolor{black}{The resulting radial velocity measurements have a mean precision of $\rm 3.58\,m/s$}

\subsection{Validation}
\label{sec:falsepos}


In this section we make use of data validation reports combined with our ground-based follow-up observations to validate the planetary nature of the three candidates orbiting TOI-282. The tests we describe here are standard procedure for the discovery of \tess planets, but we note again that for the two outer candidates the presence of anti-correlated TTVs render these steps somewhat redundant.

\subsubsection{Vetting Tests}

The \tess data validation report contains several tests to check the validity of the candidates \citep{2018PASP..130f4502T}. These tests include checking  for a difference in depth between odd and even transits, the presence of a potential secondary eclipse implying possible binarity, and any centroid offsets to ensure the event is on target. Both TOI-282.01 and .03 passed these tests successfully and were given false alarm probabilities of $2.14 \times 10^{-162}$ and $2.21 \times 10^{-34}$ respectively, indicating that these events are highly unlikely to be instrumental in nature. Having passed these tests, these candidates were ripe for follow-up observations to validate their planetary nature.

\begin{figure*}
     \centering
     \includegraphics[width = \textwidth]{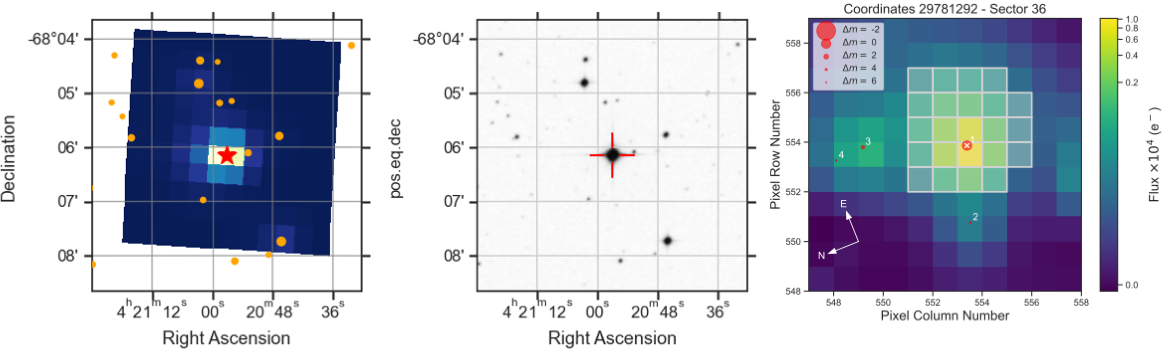}
     \caption{Images of the field of TOI-282/HD~28109, showing that the star does not have any bright contaminating sources nearby. {\bf Left Panel:} Target pixel files showing TOI-282 as a red star and other nearby sources as orange circles. The size of the circles is representative of the source's brightness. {\bf Middle Panel:} Sloan Digital Sky Survey \citep[SDSS][]{2000AJ....120.1579Y} image of the surrounding field of TOI-282 (indicated by a red cross). {\bf Right Panel:} \tess target pixel file of one of the 25 sectors in which TOI-282 was observed, created with {\sc TPFPlotter} \citep{2020AAA...635A.128A}. The pipeline aperture is shown by the white shaded region, and all nearby sources from Gaia \citep{2018AAA...616A...1G} down to a magnitude difference $\rm \Delta M =8$ are overplotted as red circles. }
     \label{fig:TPF}
 \end{figure*}

The false alarm probability for TOI-282.04 was calculated by \textsc{Sherlock} as $8 \times 10^{-5}$; \textsc{Sherlock} also provides a vetting stage for promising signals, where Field Of View (FOV) plots are generated through the internal usage of \textsc{TPFPlotter} (Figure \ref{fig:TPF}) \citep{2020AAA...635A.128A} to check for possible contamination sources; it also injects the search results into \textsc{Latte} \citep{2020JOSS....5.2101E}, which is used as the main vetting engine. \textsc{Sherlock} also provides folded lightcurves to check for even/odd transits and discard the detection of harmonics or sub-harmonics of the real signal. We scrutinise the data validation report produced by \textsc{Latte} and note no centroid offsets at the time of transits, nor were there any sharp changes in background flux.

\subsubsection{Ruling Out False Positives}


Nearby eclipsing binaries (NEB) scenarios were ruled out for all three candidates by the photometric  observations, as described in Section \ref{sec:hazel}. Subsequent observations of TOI-282.01 by \astep later further confirmed the events on target with clear detections of the transits. 

Additionally, our high-resolution observations described in Section \ref{sec:speckle} rule out any blended bound companions down to a projected orbital separation of 2.8AU. A $0.1~\rm M_{\odot}$ companion at this orbital separation would produce an RV variation of $\sim 1.6~\rm km/s$. The very low scatter in our radial velocity measurements described in Section \ref{sec:rv_intro} rules out the presence of any bound stellar-mass companions in the system.

\subsubsection{Statistical Validation with \textsc{Sherlock} and \textsc{Triceratpos}}

\textsc{Sherlock} uses the statistical validation package \textsc{Triceratpos} presented in \cite{2021AJ....161...24G} to calculate the false positive probability (FPP) for each of the candidates. \textsc{Triceratpos} queries the \tess Input Catalog (TIC) for stars within 10 pixels of the target and calculates the probability that the transits detected are caused by false positive scenarios. In all, 18 scenarios are tested, including that the signal is caused by a transiting planet on the target star with the input orbital period. \textsc{Sherlock} automates the preparation stage for \textsc{Triceratops} by preparing the lightcurves and apertures; the statistical validation is then carried out five times to produce mean probabilities for each scenario.

The criteria for statistical validation of a planetary candidate is stated as FPP < 0.015; we compute FPPs of 0.0075, 0.035 and 0.203 for TOI-282.01, .03 and .04 respectively. This immediately places TOI-282.01 well below the threshold for validation, although the probability remains too high for TOI-282.03 and .04.   

However, we are able to add further to this body of evidence by returning to the fact that this is a multi-planet system. \citet{2012ApJ...750..112L} derived an expression for a `multiplicity boost', where the presence of additional planets in the same aperture increases the probability that the signals are planetary in nature:

\begin{equation}
    P_3 \approx \frac{50\,P_1}{50\,P_1\,+\,(1-P_1)}
\end{equation}

Where $P_1$ is the probability of planethood without taking the additional planets into account, and $\rm P_3$ if the probability with the 3-planet multiplicity boost. This decreased the FPP for TOI-282.01, .03 and .04 to 0.00015, 0.0007 and 0.005 respectively. All three planets now meet the criteria for statistical validation showing that overwhelmingly, the planet hypothesis is the most favoured. We note that while the multiplicity boost described about was developed for \Kepler candidates, a similar boost for \tess candidates is described in \cite{2021ApJS..254...39G} which for small planets ($\rm R<6 R_{\oplus}$) decreases the FPP by a factor of $\approx 54$. We choose to apply the boost described in \cite{2012ApJ...750..112L} as it is more conservative. 

\subsection{Validation Conclusions}

Having ruled out all astrophysical false positive scenarios, calculated false positive probabilities and taken into account multiplicity we can now consider the system validated. 

From this point on we refer to the three validated planets by their formal names in order of increasing semi-major axis:

\begin{itemize}
    \item TOI-282.04 $\rightarrow$ HD~28109~b
    
    \item TOI-282.01 $\rightarrow$ HD~28109~c
    
    \item TOI-282.03 $\rightarrow$ HD~28109~d
\end{itemize}

\section{Analysis}
\label{sec:analysis}
With the planetary nature of the signals validated, we now present our analysis of all available data for HD~28109~b, c and d. We begin by describing our global analysis of the available photometry for the system in Section \ref{sec:glob_phot}, followed by our dynamical analysis in Section \ref{sec:dynamic}. In Section \ref{sec:rv_phot} we make use of priors from our photometric and dynamical fits to fit the available radial velocity data. 


\subsection{Global Photometric Analysis}
\label{sec:glob_phot}

We carried out a global photometric analysis of the datasets described in Section \ref{sec:validation} using \textsc{Allesfitter} \citep{2021ApJS..254...13G, 2019ascl.soft03003G}. \textsc{Allesfitter} is a publicly available versatile inference package capable of jointly fitting photometric and radial velocity datasets from different instruments using \textsc{ellc} \citep{2016AAA...591A.111M} for lightcurve and RV models, \textsc{emcee} \citep{2013PASP..125..306F} for MCMC, \textsc{Dynesty} \citep{2020MNRAS.493.3132S} for dynamic nested sampling, and \textsc{celerite} \citep{2017ascl.soft09008F} for Gaussian Process (GP) models. We chose to use this package as it can simultaneously fit for white (uncorrelated) noise and red (correlated) noise for different instruments, while offering a wide selection of models to choose from and compare evidence.   

We adopted the signal parameters found in Section \ref{sec:cand_search} as uniform priors, and the stellar parameters derived in Section \ref{sec:starchar} as Gaussian priors and we fit using the nested sampling algorithm. Nested sampling \citep{2004AIPC..735..395S} works by drawing a number of live points from the prior distribution, and then removing the point with the lowest likelihood. Another point is then drawn from the prior while requiring that the overall likelihood now be higher than before. This process then repeats until the change in Bayesian evidence, parametrised as $\Delta \ln Z$, falls below a certain threshold. We begin the algorithm with 1500 live points and set the threshold to $\Delta \ln Z < 0.01$ as recommended by the \textsc{Allesfitter} documentation.  

We computed quadratic limb darkening coefficients $u_1$ and $u_2$ for HD~28109 in \tess and {\it $R_c$} filters using \textsc{Pyldtk} \citep{2015MNRAS.453.3821P} and the PHOENIX stellar atmosphere library provided by \citep{2013AAA...553A...6H}. The resulting coefficients are presented in Table \ref{tab:glob_fit}; these are also input to our global fit as Gaussian priors.

We initially fit for a two-planet model using only HD~28109~c and HD~28109~d, and then for a three-planet model incorporating HD~28109~b. One of the key advantages of nested sampling over MCMC (Markov Chain Monte Carlo) is the calculation of the Bayesian evidence at each stage of the algorithm; this allows us to determine which model is statistically favoured by calculating the Bayes' factor \citep{doi:10.1080/01621459.1995.10476572}. In both cases, we fit simultaneously for all transit parameters, quadratic stellar limb-darkening coefficients in the \cite{2013MNRAS.435.2152K} parametrisation, as well as the baseline model and the white noise scaling for each instrument. These latter two ensure that uncertainties arising from instrument systematics are appropriately propagated to derived physical parameters. 

In Section \ref{sec:cand_search} we found that all three planets had the highest SNR in \tess when the flux was detrended; to detrend the data for semi-periodic stellar variability (red noise) present in all instruments, we simultaneously fit a GP using a Simple Harmonic Oscillating (SHO) kernel. The parametrisation used in the fit is summarised in Table \ref{tab:glob_fit}.

We find that the 3-planet model is vastly statistically favoured over the 2-planet model, with a Bayes Factor in excess of 60\,000. A \rm $BF > 150$ is usually considered strong enough statistical evidence to confirm a more complex model \citep{doi:10.1080/01621459.1995.10476572}. The fit's result further confirms the existence of HD~28109~b in our data. 

In both the 2- and 3-planet model fits, eccentricity and argument of periastron were set as free parameters. We note that the resulting derived orbital parameters \textcolor{black}{suggest} that the two outer planets \textcolor{black}{could} have small \textcolor{black}{to moderate} eccentricities ($e_{\rm c} = 0.116_{-0.079}^{+0.30}$ and $e_{\rm d} = 0.19_{-0.15}^{+0.34}$) and \textcolor{black}{yield calculated} host densities within $1\sigma$ of the prior, while the inner planet yields a very large eccentricity of \textcolor{black}{$e_{\rm b} = 0.809_{-0.083}^{+0.049}$} and a \textcolor{black}{calculated} host density 5-$\sigma$ \textcolor{black}{below} the prior. The host density is \textcolor{black}{calculated from the transit parameters of each planet individually} by \textsc{Allesfitter} using the expression presented in \cite{2003ApJ...585.1038S}:

\begin{equation}
    \rho_{\star ,\rm obs} \equiv \frac{3 \pi (a/R_\star)^3_{obs}}{GP^2}
\end{equation}

where $\rho_{\star , \rm obs}$ is the host density derived from transit parameters assuming a Keplerian circular orbit. The manner and extent of the discrepancy between  $\rho_{\star , obs}$ and the external prior on stellar density can be indicative of erroneous assumptions made regarding the orbits \citep{2014MNRAS.440.2164K}. In particular, for such an eccentric orbit we would expect the ratio $\rho_{obs} / \rho_{true}$ to be much greater than unity. 

We therefore repeat the 2- and 3-planet fits constraining eccentricities to zero and then compare the evidence for each of the models to see which is statistically most likely. We find that all 3-planet models are preferred over all 2-planet models, with Bayes Factors in excess of 50\,000. We also find that all circular models are preferred over models with non-zero eccentricities. In the case of the 3-planet circular model vs. the 3-planet eccentric model we find a Bayes Factor of almost 3\,000, \textcolor{black}{indicating that a more complex lightcurve modelling with eccentric orbits is not preferred.}



\begin{figure}
    \centering
    \includegraphics[width =\columnwidth]{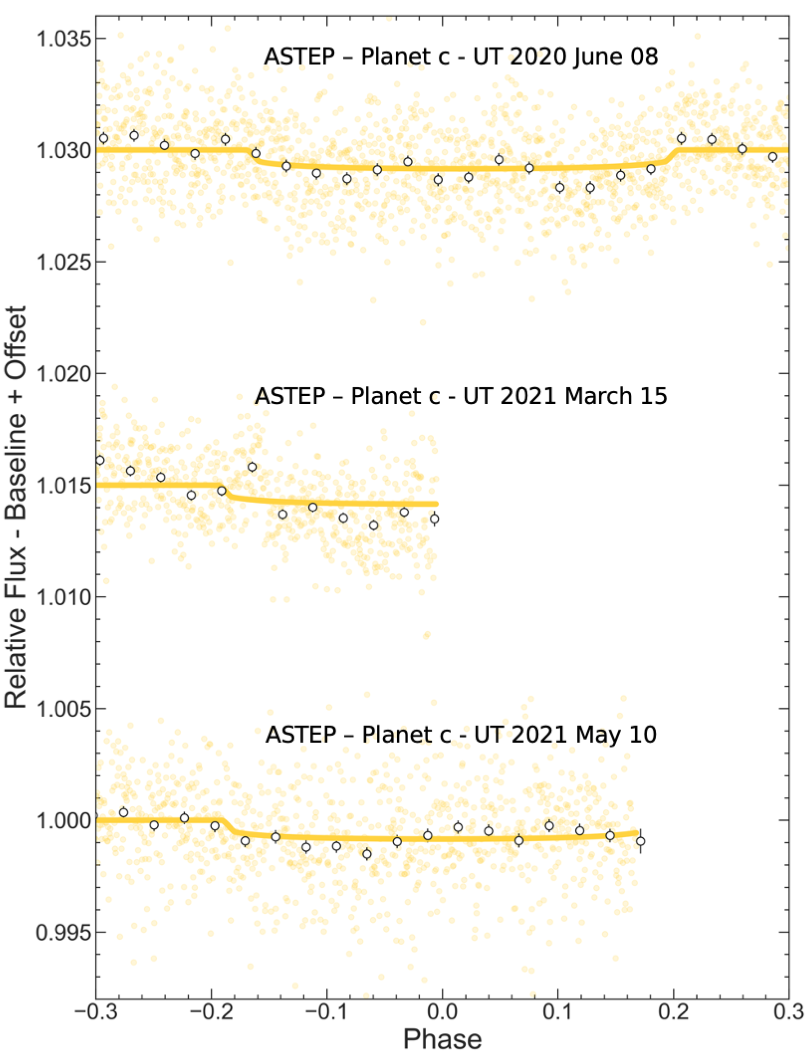}
    \caption{Detrended transits observed with \astep on the nights of 2020 June 08, 2021 March 15 and 2021 May 10. The green points are the detrended flux, while the white circles are binned. Curves superimposed are the best-fitting models for each transit. }
    \label{fig:astep_trans}
\end{figure}

\begin{figure}
    \centering
    \includegraphics[width =\columnwidth]{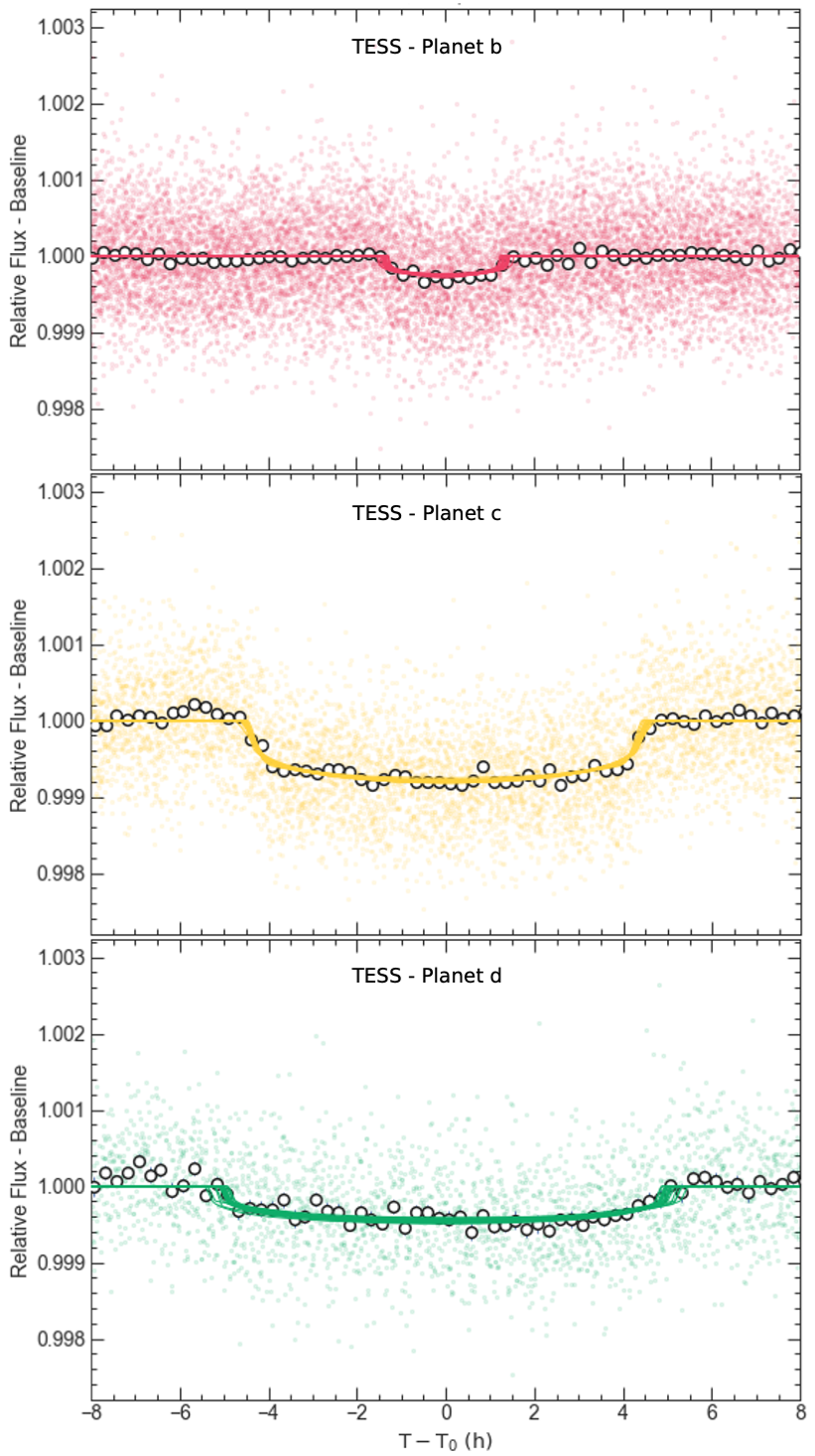}
    \caption{Global fit of the \tess photometric data for each of the planets of HD~28109. Points in the background are the phase-folded 2-minute cadence data, while white circles are the binned points. Curves are 20 samples drawn from the posterior.}
    \label{fig:tess_trans}
\end{figure}

The results of the global fit, including derived parameters, are presented in Table \ref{tab:glob_fit}. In Figure \ref{fig:astep_trans} we present the three individual \astep transits for planet c along with the best fitting model, and in Figure \ref{fig:tess_trans} we present the \tess phase folded lightcurves for each planet along with the best fitting model.

\begin{table*}
\centering
\caption{Priors used in our fit, along with fitted and derived parameters. Uniform priors are indicated as $\rm \mathcal{U}(lower~bound,~upper~bound)$ and normal priors are indicated as $\rm \mathcal{N}(mean,~standard~deviation)$. }
\resizebox{\textwidth}{!}{%
\begin{tabular}{@{}ccccc@{}}
\toprule
{ \bf }                      & { \bf HD~28109~b}                       & { \bf HD~28109~c}                          & { \bf HD~28109~d} & \textcolor{black}{\textbf{Source}}      \\ \midrule \midrule
\multicolumn{5}{c}{\textbf{Fit Parametrisation and Priors}}                           \\ \midrule
Transit Depth $\rm R_p / R_\star$         & $\mathcal{U}(0.008, 0.015)$   & $\mathcal{U}(0.023, 0.035)$   &  $\mathcal{U}(0.017, 0.037)$         \\
\textcolor{black}{Inverse Scaled Semi-major Axis} $\rm (R_\star + R_p) / a$   & $\mathcal{U}(0.05, 0.06)$    & $\mathcal{U}(0.015, 0.035)$   &  $\mathcal{U}(0.015, 0.03)$        \\
Orbital Inclination $\rm \cos{i}$               & $\mathcal{U}(0.000, 0.04)$    & $\mathcal{U}(0.000, 0.04)$   &  $\mathcal{U}(0.000, 0.04)$       \\
Transit Epoch $\rm T_{0}$   & $\mathcal{U}(2458344.7, 2458344.9)$   & $\mathcal{U}(2458337.12, 2458337.32)$   &  $\mathcal{U}(2458355.58, 2458355.98)$  \\
Planet Mass $\rm M_p / M_{\star}$ & $\mathcal{U}(0.0, 0.000026)$  & $\mathcal{U}(0.0, 0.0005)$ & $\mathcal{U}(0.0, 0.0005)$  \\
Period $\rm P$           & $\mathcal{U}(22.8, 23.0)$    & $\mathcal{U}(55.6, 56.6)$   &  $\mathcal{U}(84.2, 86.0)$       \\
Eccentricity $\rm e $ & $\mathcal{U}(0.0, 0.5)$    & $\mathcal{U}(0.0, 0.5)$   &  $\mathcal{U}(0.0, 0.5)$       \\
Longitude of Periapsis $\varpi$ & $\mathcal{U}(0, 360)$    & $\mathcal{U}(0, 360)$   &  $\mathcal{U}(0, 360)$        \\ \midrule
\multicolumn{5}{c}{\textbf{Limb Darkening Coefficients}} \\ \midrule
\textcolor{black}{\tess $\rm u_1$} & \textcolor{black}{$0.4041\pm0.0007$} & \textcolor{black}{\astep $\rm u_1$} & \textcolor{black}{$0.4758\pm0.0009$} \\
\textcolor{black}{\tess $\rm u_2$} & \textcolor{black}{$0.1324\pm0.0018$} & \textcolor{black}{\astep $\rm u_2$} & \textcolor{black}{$0.1392\pm0.0022$}  \\
\textcolor{black}{\tess $\rm q_1$} & \textcolor{black}{$\mathcal{N}(0.2878, 0.0024)$} & \textcolor{black}{\astep $\rm q_1$} & \textcolor{black}{$\mathcal{N}(0.3782, 0.0034)$}  \\
\textcolor{black}{\tess $\rm q_2$} & \textcolor{black}{$\mathcal{N}(0.3766, 0.0013)$} & \textcolor{black}{\astep $\rm q_2$} & \textcolor{black}{$\mathcal{N}(0.3868, 0.0014)$} \\\midrule
\multicolumn{2}{c}{\textbf{External Priors}}  & \multicolumn{2}{c}{\textbf{GP Priors}}                              \\\midrule
Stellar Mass $\rm M_{\star}$  & $\mathcal{N}(1.26, 0.08)$ & Amplitude Scale $\mathrm{gp \ln S_0 (flux)}$ & $\mathcal{U}(-15, 5)$      \\
Stellar Radius  $\rm R_{\star}$   & $\mathcal{N}(1.446, 0.035)$  & Damping $\mathrm{gp \ln Q (flux)}$   & $\mathcal{U}(-10, 10)$  \\
Stellar Effective Temperature $\rm T_{\rm eff}$  & $\mathcal{N}(6120, 50)$ & Frequency $\mathrm{gp \ln \omega_0 (flux)}$ & $\mathcal{U}(-10, 10)$            \\\midrule
\multicolumn{5}{c}{{ \bf Fitted Parameters}}                        \\ \midrule
$\rm R_p / R_\star $      & $0.01441_{-0.00037}^{+0.00031}$ &    $0.02632\pm0.00027$           & $0.01999_{-0.00037}^{+0.00035}$    & \textcolor{black}{Fixed Lin. Ephem. Fit}     \\
$\rm (R_\star + R_p) / a$         & $0.05017_{-0.00012}^{+0.00022}$ & $0.02231_{-0.00066}^{+0.00073}$    &  $0.01688_{-0.00047}^{+0.00053}$ & \textcolor{black}{Fixed Lin. Ephem. Fit}  \\
$\rm \cos{i}$       & $0.03964_{-0.00040}^{+0.00024}$ & $0.0081_{-0.0021}^{+0.0018}$       & $0.0064_{-0.0015}^{+0.0013}$     & \textcolor{black}{Fixed Lin. Ephem. Fit}     \\
$\rm T_{0}$ $(\mathrm{BJD})$    & $2458344.81772\pm0.00757$ & $2458377.80109_{-0.00733}^{+0.00724}$   & $2458355.67324\pm0.00432$  & \textcolor{black}{Dynamical Fit} \\
$\rm P$ $(\mathrm{d})$       & $22.89104_{-0.00036}^{+0.00035}$ &  $56.00819_{-0.00202}^{+0.00194}$       & $84.25999_{-0.00662}^{+0.00744}$  & \textcolor{black}{Dynamical Fit}   \\
$\rm e$ &  $< 0.3307$   &  $< 0.1203$ & $< 0.0864$ & \textcolor{black}{Dynamical Fit} \\ \midrule
\multicolumn{2}{c}{\textcolor{black}{ \bf Fitted RV Jitter}} & \multicolumn{2}{c}{\textcolor{black}{ \bf Fitted GP Hyperparameters}} &   \\ \midrule
\textcolor{black}{$\ln{\sigma_\mathrm{jitter; HARPS}}$ ($\ln{ \mathrm{km/s} }$)} &  \textcolor{black}{$-7.8_{-2.8}^{+1.6}$ }   &\textcolor{black}{$\mathrm{gp \ln S_0 (flux)}$} &  \textcolor{black}{$-19.72_{-0.19}^{+0.20}$}  & \textcolor{black}{Fixed Lin. Ephem. Fit} \\
\textcolor{black}{$\ln{\sigma_\mathrm{jitter; espresso}}$ ($\ln{ \mathrm{km/s} }$)} &  \textcolor{black}{$-5.99_{-0.36}^{+0.38}$}  &\textcolor{black}{$\mathrm{gp \ln Q (flux)}$} & \textcolor{black}{$-4.9_{-2.2}^{+1.9}$} & \textcolor{black}{Fixed Lin. Ephem. Fit}\\
 & & \textcolor{black}{$\mathrm{gp \ln \omega_0 (flux)}$} & \textcolor{black}{$6.0_{-1.9}^{+2.2}$}   & \textcolor{black}{Fixed Lin. Ephem. Fit}  \\\midrule
\multicolumn{5}{c}{{ \bf Derived Parameters}}        \\ \midrule
Companion radius; $\rm R_\mathrm{p}$ ($\mathrm{R_{\oplus}}$)               & \textcolor{black}{$2.199_{-0.10}^{+0.098}$}    &  \textcolor{black}{$4.23\pm0.11$}   &  \textcolor{black}{$3.25\pm0.11$} & \textcolor{black}{Fixed Lin. Ephem. Fit}  \\
Semi-major axis; $\rm a$ (AU)                                              & $0.1357\pm0.0034$               & $0.308\pm0.011$                    & $0.411\pm0.016$  & \textcolor{black}{Fixed Lin. Ephem. Fit}   \\
Inclination; $\rm i$ (deg)                                                 & $87.725_{-0.012}^{+0.023}$     & $89.543_{-0.086}^{+0.093}$            & $89.682_{-0.082}^{+0.093}$   & \textcolor{black}{Fixed Lin. Ephem. Fit}   \\
Impact parameter; $\rm b$                                                  & $0.8007_{-0.0084}^{+0.0055}$    & $0.365_{-0.068}^{+0.058}$          & $0.339_{-0.093}^{+0.075}$   & \textcolor{black}{Fixed Lin. Ephem. Fit}  \\
Total transit duration; $\rm T_\mathrm{tot}$ (h)                           & $5.392_{-0.073}^{+0.10}$        & $8.973\pm0.060$                    & $10.13_{-0.11}^{+0.14}$  & \textcolor{black}{Fixed Lin. Ephem. Fit}  \\
Full-transit duration; $\rm T_\mathrm{full}$ (h)                           & $4.997_{-0.081}^{+0.11}$        & $8.425\pm0.070$                    & $9.66_{-0.12}^{+0.14}$     & \textcolor{black}{Fixed Lin. Ephem. Fit}  \\
Equilibrium temperature; $\rm T_\mathrm{eq}$ (K)                           & $881.2\pm7.5$                   & $585.2_{-8.3}^{+8.7}$               & $506.5_{-8.0}^{+8.7}$ & \textcolor{black}{Fixed Lin. Ephem. Fit}  \\
Transit depth \tess; $\rm \delta_\mathrm{tr; TESS}$ (ppt)   &    \textcolor{black}{$0.188_{-0.016}^{+0.013}$ }  & \textcolor{black}{$0.834_{-0.017}^{+0.015}$} &  \textcolor{black}{$0.489_{-0.024}^{+0.021}$}    & \textcolor{black}{Fixed Lin. Ephem. Fit}      \\
Transit depth \astep; $\rm \delta_\mathrm{tr; ASTEP}$ (ppt) &             -             & \textcolor{black}{$0.854_{-0.018}^{+0.016}$}           &        -     & \textcolor{black}{Fixed Lin. Ephem. Fit}   \\
Companion TTV mass; $\rm M_\mathrm{p; TTV}$ ($\mathrm{M_{\oplus}}$)  &           -            &   $7.943_{-3.046}^{+4.227}$     &     $5.681_{-2.110}^{+2.738}$  & \textcolor{black}{Dynamical Fit}  \\
Companion RV mass; $\rm M_\mathrm{p; RV}$ ($\mathrm{M_{\oplus}}$)  &    $18.496_{-7.609}^{+9.120}$    & - & - & \textcolor{black}{RV Fit}      \\ 

\end{tabular}%
}
\label{tab:glob_fit}
\end{table*}

\subsection{Dynamical Analysis}
\label{sec:dynamic}


In this section we consider a non-Keplerian model for the planets of HD~28109; we first fit individual transit times and then use these timings to infer dynamical masses for the planets.


\subsubsection{Transit Timing Variations}

Given the proximity of the two outer planets to a first order 3:2 mean-motion resonance (within $\sim 0.3\%$), we would expect these planets to experience some mutual gravitational influence leading to transit timing variations (TTVs). This is supported by visual inspection of the individual transit fits resulting from the nested sampling, as several appear to have noticeable offsets in time between the model and the data. 


We therefore re-fit the photometry once again using \textsc{Allesfitter}, this time allowing the midtime of each individual transit to vary about a linear ephemeris for all three planets. 

When allowing for TTVs, we fit for all the same parameters as in the linear fit, except that we now fix the \textcolor{black}{linear ephemeris} using the values derived from the most statistically favoured fit in Section \ref{sec:glob_phot}. We also constrain eccentricities to zero for all three planets and once again apply Gaussian priors on the stellar density and quadratic limb darkening coefficients.

In preparation for a TTV fit, \textsc{Allesfitter} attempts to guess the location of the flux minimum during each transit present in the data, with these guesses then used as uniform priors for the fit. However, we find that almost all of the guesses for planet b are affected by local stellar variability or nearby transits of other planets biasing the fit; instead, for each transit's mid-time, we impose a uniform prior defined as the linear predictions $\pm 120$ minutes for all planets. Once again we simultaneously fit a GP with a SHO kernel to account for stellar variability. 

\begin{figure}
    \includegraphics[width=\columnwidth]{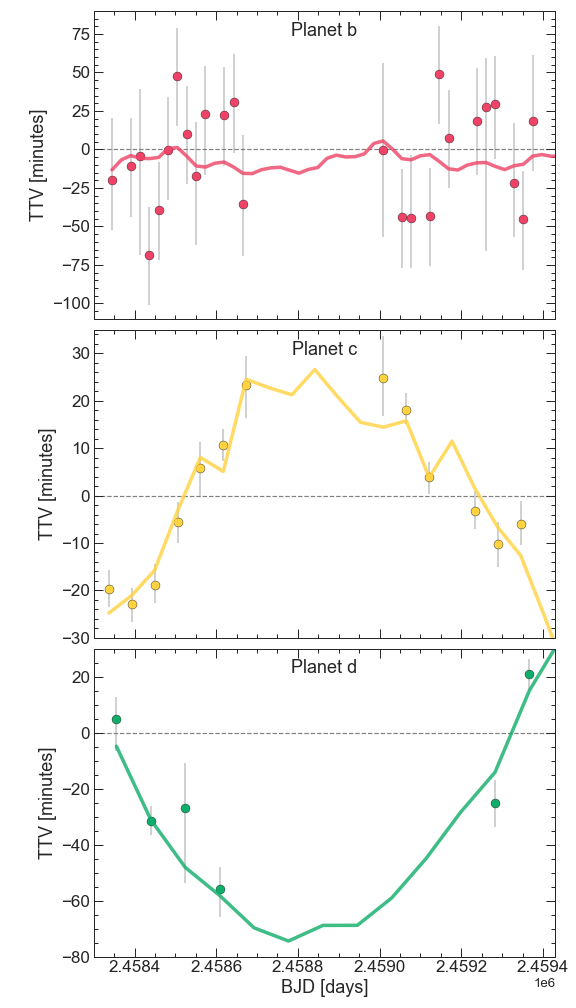}
    \caption{Transit timing variations (in minutes) for each of the planets of HD~28109 spanning two years of \tess data plus ground-based photometric monitoring of planet c. The points show the difference between the fitted midtimes and the linear predictions, while the lines show the model predictions from our dynamical analysis in Section~\ref{sec:nesvorny}.}
    \label{fig:all_ttvs}
\end{figure}

The results of this fit are presented in Figure \ref{fig:all_ttvs}. HD~28109~c and d present very clear and significant anti-correlated TTVs with peak to peak amplitudes of $\sim 50~ \rm mins$ and $\sim 100~ \rm mins$ respectively. While the shifts in timing do appear to show a sinusoidal variation between early and late, it is most likely that what we see here is the shorter term `chopping' part of the signal \citep{2015ApJ...802..116D} given that the `Super-Period' for these planets is expected to be in excess of 9\,000 days. 

For HD~28109~b the errors on the timings are very large due to how shallow individual events are \textcolor{black}{($\rm \approx 0.19ppt$)} and therefore our results are consistent with no TTVs. We note that for some individual events the errors were significantly smaller than average. Visual inspection of each transit fit showed that this is probably caused by artifacts in the lightcurves. We therefore amplified the errors on all timings of planet b to be at least the mean timing uncertainty. 

\subsubsection{Mass Estimates From TTVs}
\label{sec:nesvorny}


We perform dynamical analysis of the observed transit times with a symplectic $N$-body integrator code described in \cite{2012Sci...336.1133N}. We instruct the code to simultaneously fit all measured transit times, using MULTINEST \citep{2008MNRAS.384..449F} to perform the regression, setting the integration time-step to 0.5 d, roughly 1/50 of the planet b's orbital period. We consider dynamical models with 2, 3 and 4 planets. The model with 3 planets has 20 parameters: planet-to-star mass ratios, orbital periods, eccentricities, longitudes of periapsis, impact parameters, difference in nodal longitudes, and reference epochs between a reference time (2458337 BJD UTC) and the first observed transit of each planet. \textcolor{black}{The nodal longitude of the innermost planet in each fit is fixed at $270^{\circ}$, and therefore} the 2 and 4 planet models have 13 and 27 parameters, respectively. We use uniform priors for all parameters except for the impact parameter, since our light-curve fits provide strong constraints which are used here. The host star mass was fixed at $M_\star=1.26~\rm M_\odot$. The uncertainties discussed below therefore do not account for the uncertainty in the stellar mass.

We find that the existing measurements are not good enough to uniquely determine the planet properties. The first problem arises because there is a large radial separation between the inner planet b and the two outer planets: the measured TTVs of c and d thus most likely reflect their mutual interaction (rather than those of planet b). In addition, given the relatively large measurement errors of b's TTVs it is not clear whether b's TTVs contain any useful information about planets c and d. Thus, as b can be decoupled from the rest of the system it is difficult to meaningfully constrain its properties. Indeed, when we instruct MULTINEST to perform a 3-planet fit, the mass of planet b always reaches to the upper edge of the mass prior even if this is clearly nonphysical (i.e., mass exceeding that of Jupiter for the estimated radius \textcolor{black}{$R_b=2.199$ $R_\oplus$)}. We therefore limit the uniform prior on the mass of planet b to $\rm M_{\rm b} = 11 M_{\oplus}$.

All fits performed here confirm that the masses of c and d are in the planetary range. Our best fit indicates $\rm m_{\rm c}/\rm M_*=(1.9_{-0.7}^{+1.0}) \times 10^{-5}$ and $\rm m_{\rm d}/\rm M_*=(1.4_{-0.5}^{+0.6}) \times 10^{-5}$, 
suggesting that both planets have low densities. The orbital eccentricities of c and d are most likely low ($\rm e<0.1$), but some 2-planet fits identify modes with eccentric orbits ($\rm e>0.1$). In Table \ref{tab:glob_fit} we present the $2 \sigma$ upper limits for the eccentricities of all three planets as this parameter remains poorly constrained; the mean values of the posterior distributions for the eccentricities are $\rm e_b = 0.1519_{-0.0944}^{+0.0894}$, $\rm e_c =0.0391_{-0.0231}^{+0.0406}$, and  $\rm e_d =0.0238_{-0.0157}^{+0.0313}$. The posteriors for the longitude of periastron are not included in Table \ref{tab:glob_fit} as they are unconstrained by our fits. 
 
Given the wide orbital separation between planets b and c, we perform an additional fit including a putative fourth planet. This planet may be non-transiting or too small to be detected. We used a uniform prior between 28 d to 44 d for the 4th planet orbital period. The 4-planet fits give masses of c and d that are consistent with the masses reported above (and low orbital eccentricities). The best parameters for the hypothetical 4th planet are $\rm m_4/M_* \simeq 7.6 \times 10^{-6}$ and $\rm P_4 \simeq 39.3$ d. With this orbit, the two inner planets would be wide of the 3:2 resonance. The Bayesian evidence for this fit is significantly lower than for the 3-planet fit, indicating that the 3-planet fit is preferred.

\begin{figure}
    \centering
    \includegraphics[width=\columnwidth]{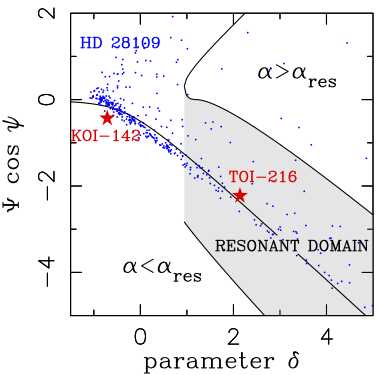}
    \caption{The 3:2 resonance structure diagram following \citet{2016ApJ...823...72N}. Systems with $\alpha=a_{\rm c}/a_{\rm d} < \alpha_{\rm res}$ ($\alpha > \alpha_{\rm res}$), where $\alpha = 0.763$ corresponds to the exact resonance, have orbits just wide (narrow) of the resonance. The resonance region where $\psi$ librates is shaded. The separatrices and stable point are solid. Three planetary systems are plotted: HD 28109 (this work), TOI-216  \citep{2021AJ....161..161D, 2021arXiv211009577N} and KOI-142 \citep{2013ApJ...777....3N}. TOI-216 is firmly in the resonance and KOI-142 is firmly outside the resonance. For HD~28109, the blue dots show a sample of 500 posterior solutions from our dynamical 3-planet fit, of which roughly two-thirds are non-resonant and roughly one-third are resonant.  }
    \label{fig:resonance}
\end{figure}

The eccentricities of planets c and d are low and when that is the case, the resonant and near-resonant dynamics can be studied analytically following \cite{2016ApJ...823...72N}. There are three variables to consider, each of them being a combination of orbital elements. Constant $\delta$ is an orbital invariant that defines the position of the system relative to the 3:2 resonance, the resonant angle $\psi$ is a function of the usual 3:2 resonant angles $\sigma_1$ and $\sigma_2$, and variable $\Psi$ is a combination of planetary masses, semi-major axes and eccentricities (see \cite{2016ApJ...823...72N} for details). The resonant librations of $\psi$ can only happen for $\delta > 0.945$. Figure \ref{fig:resonance} shows position of planets c and d in the terms of the dynamical variables $\delta$, $\psi$, and $\Psi$. About two thirds of the posterior samples are non-resonant with the two orbits being slightly wide of the exact resonance. About one third of the sample, however, show resonant librations with a generally low libration amplitude. Additional observations will be needed to better constrain the location of the c and d planets relative to the 3:2 resonance. 

\begin{figure}
    \centering
    \includegraphics[width=\columnwidth]{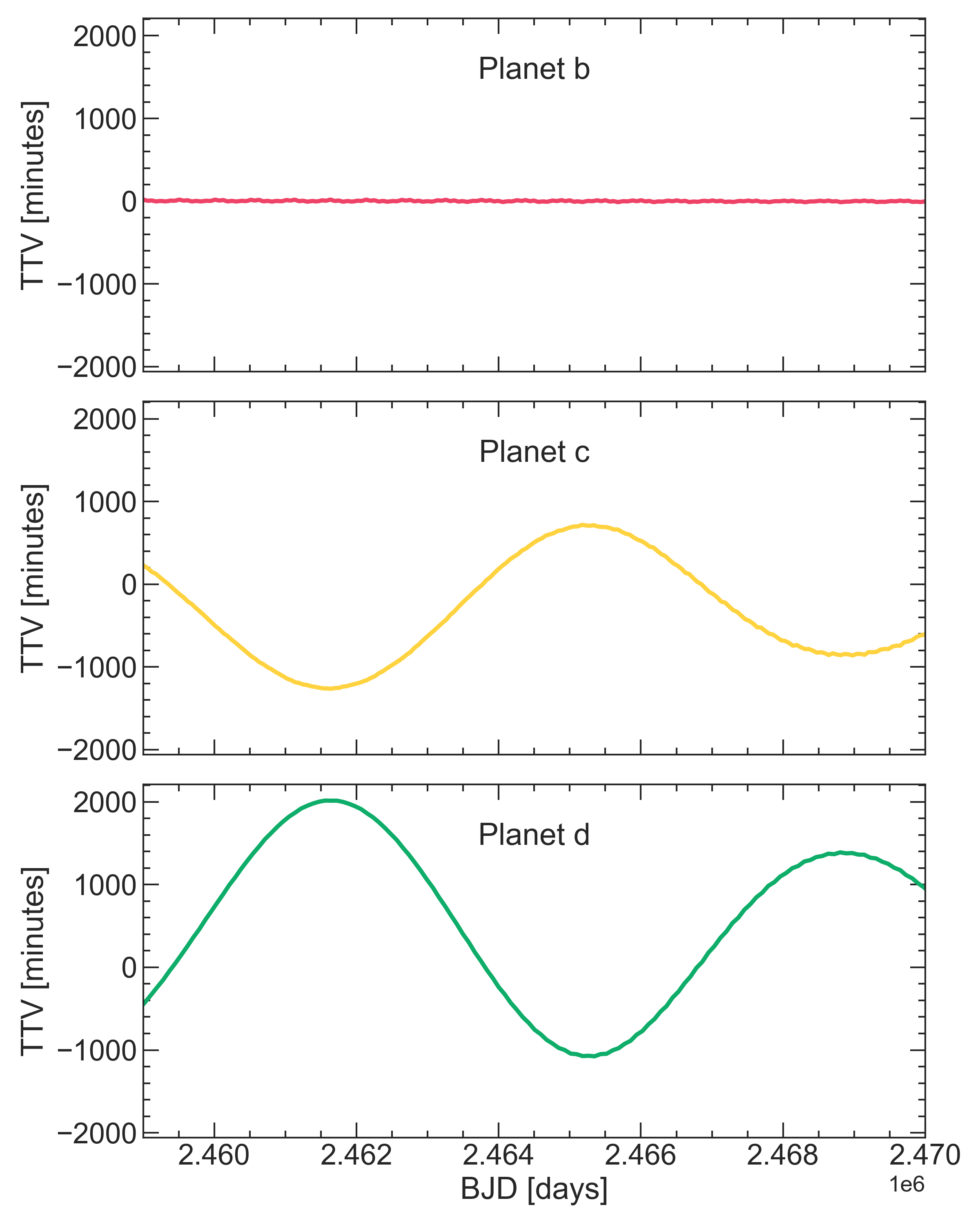}
    \caption{Model predictions of the transit timings variations for the planets of HD~28109 for the next $\rm 11\,000$ days, spanning more than one full TTV period for the two outer planets.}
    \label{fig:ttv_pred}
\end{figure}

We predict that the amplitudes of the TTVs will be very large ($\rm \sim 1000\,min$ for planet c and $\rm \sim 2000\, min$ for planet d) once one whole TTV period has been sampled. In Figure \ref{fig:ttv_pred} we present the predicted TTVs of the system for the next $\rm 11\,000\, days$. Continued monitoring of the system will allow us to both improve timing predictions and refine the dynamical mass measurements of planets c and d. 

\subsection{Radial Velocity Analysis }
\label{sec:rv_phot}

The global photometric analysis yielded strong priors on all orbital parameters; additionally, we can place priors on the masses of the two outer planets from the dynamical fit. We therefore fit the radial velocity data set to gain an estimate on the mass of the innermost planet using \textsc{Allesfitter}. We adopt Gaussian priors on the all orbital parameters for the three planets ($\rm P,~T_0,~,R_p/R_*,~(R_* + R_p) / a,~cos{i},$), as well as the radial velocity semi-amplitudes of the outer planets ($\rm K$). We place a wide uniform prior between $0.1 - 6.0~\rm m/s$ (corresponding to $0.5 - 30~ \rm M_{\oplus}$) on the semi-amplitude of the inner planet and fit for all three periods, reference times and semi-amplitudes. \textcolor{black}{In our models, we vary the RV offset and RV jitter for each individual dataset. The jitter parameter appears as an additional RV noise term added in quadrature to the nominal RV uncertainties and subsums any instrumental or stellar effects (such as activity) not considered by the formal RV errors.} 

\begin{figure}
    \centering
     \includegraphics[width=\columnwidth]{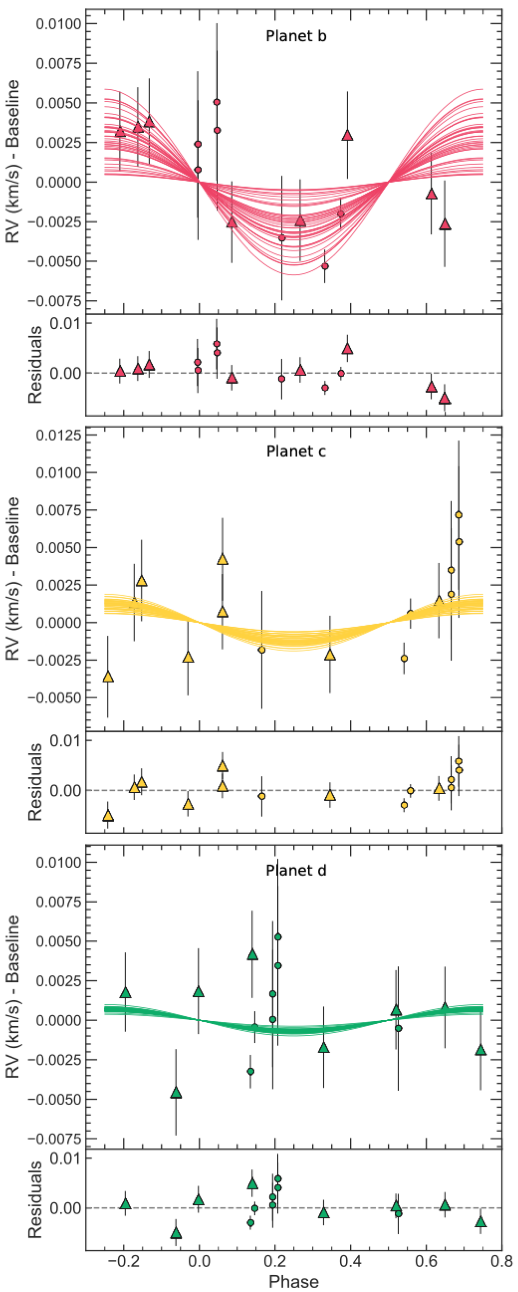}    
     \caption{Phased radial velocity measurements for each of the three planets. Triangles are points from ESPRESSO, while circles are points from HARPS; the curves are 20 model samples from the posterior. \textcolor{black}{The data uncertainties include the RV jitter.}
}
     \label{fig:RV}
 \end{figure}

In Figure \ref{fig:RV} we present the phased radial velocity curves. The radial velocity semi-amplitude posterior for the inner planet yields a mass estimate of $\rm M_{\rm b}=18.50_{-7.61}^{+9.12}~\rm M_{\oplus}$; \textcolor{black}{for the outer planets we find mass estimates consistent with the estimates from the dynamical analysis: $\rm M_{\rm c}=7.99_{-2.03}^{+2.13}~\rm M_{\oplus}$ and $\rm M_{\rm d}=5.53_{-0.94}^{+0.98}~\rm M_{\oplus}$}. We do however note that to precisely constrain the masses of the system far more measurements are needed, and therefore this result is just a first estimate. 

\section{System Discussion}
\label{sec: architecture}

\begin{figure}
    \centering
    \includegraphics[width = \columnwidth]
    {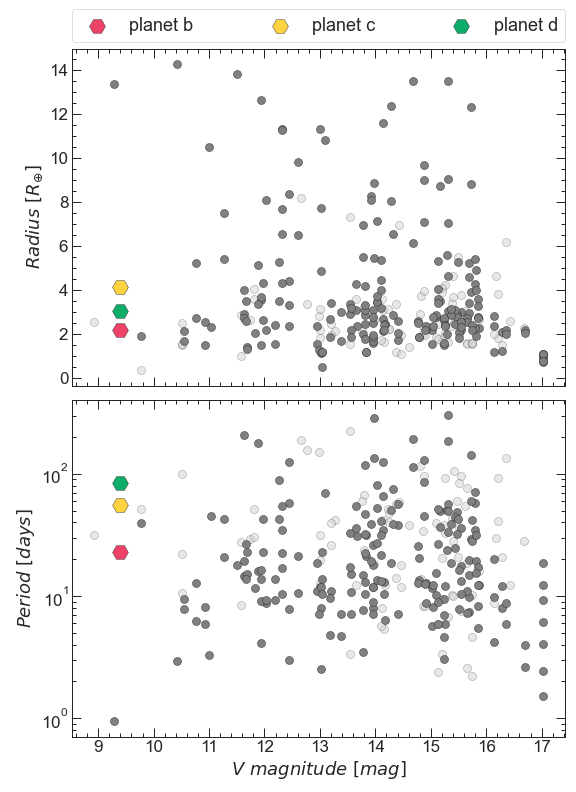}
    \caption{Distribution of known multi-planetary systems exhibiting transit timing variations. The planets of HD~28109 are shown as hexagons in violet (planet b), green (planet c) and pink (planet d); other planets are either shown in dark grey if the have a measured mass (by RV or TTV) or light grey if they do not. In the upper panel we show the distribution in planet radius according to host star {\it V} mag,  while in the bottom panel we show the distribution in orbital period.}
    \label{fig:oth_ttv}
\end{figure}

HD~28109 is host to at least three planets, with two very close to a first order mean-motion resonance. So far, this is the brightest \tess star, and the third brightest star overall, known to host TTV planets. In Figure \ref{fig:oth_ttv} we present the current sample of planets known to show transit timing variations in multi-planetary systems, plotted as planet radius and orbital period vs.~host V magnitude, highlighting the position of the three planets of the HD~28109 system\footnote{Data for this plot was retrieved from the NASA Exoplanet Archive \citep{2013PASP..125..989A} on 2021 October 20: \url{https://exoplanetarchive.ipac.caltech.edu/cgi-bin/TblView/nph-tblView?app=ExoTbls\&config=PSCompPars}.}. Planets plotted as dark grey circles have measured masses (by RV or TTV) while grey circles have masses estimated by mass-radius relations. The only host brighter than HD~28109 known to host TTV planets with measured masses is WASP-18, but as the outer planet is non-transiting full characterisation of the system is not yet possible \citep[][]{2019AJ....158..243P}.

In this section we discuss the system in the context of other known exoplanetary systems, and current and near-future instrumentation capabilities. 

\subsection{The planets of HD~28109 on the Mass-Radius Diagram}

\begin{figure*}
    \centering
    \includegraphics[width = 0.75\textwidth]
    {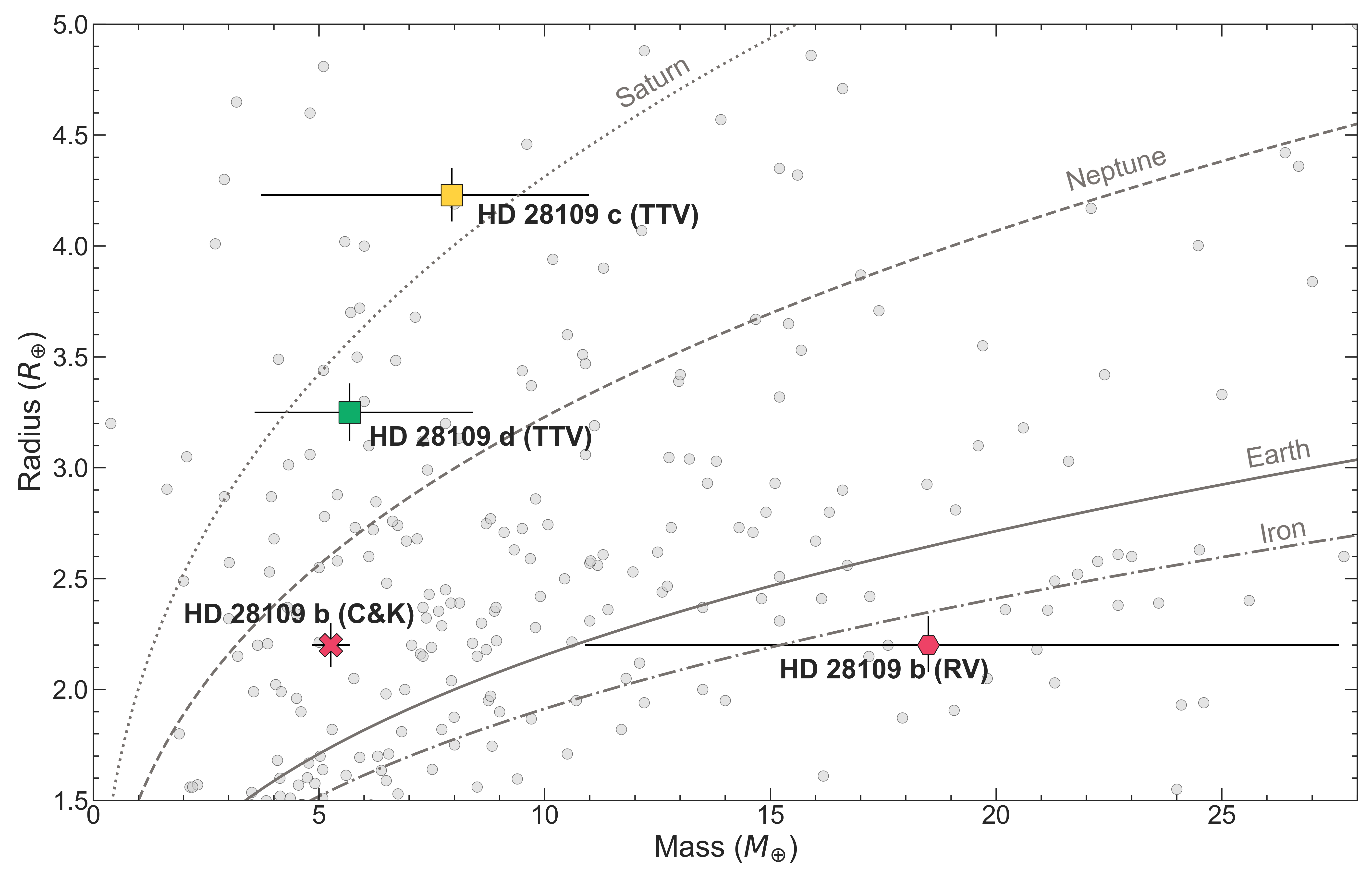}
    \caption{Mass-radius diagram showing the positions of HD~28109~b, c and d. Planets c and d are plotted as squares with their masses are estimated from TTVs. As the mass of planet b is still poorly constrained, we plot its position using its RV estimated mass as a hexagon, while the average mass for planets of its radius (estimated mass following \citet{2017ApJ...834...17C}) is plotted as a cross.}
    \label{fig:mass_rad}
\end{figure*}


In Figure \ref{fig:mass_rad} we present a Mass-Radius diagram showing the positions of HD~28109~b, c, and d in relation to other known exoplanets with masses measured using TTVs or precise radial velocities. The mean densities of Earth, Neptune, Saturn and pure iron are also shown. 

The masses of HD~28109~c and d estimated from TTVs indicate that they are under-dense compared with most other sub-Neptunian mass planets. Given the paucity of planets in this size and mass regime with precisely constrained parameters at orbital separations comparable to HD~28109 b and c, we cannot determine at this time whether or not their densities are truly anomalous.


The mass estimate of HD~28109~b is rather more puzzling. First estimates from radial velocities in this work suggest that this planet is significantly denser than typical planets of its size.
Atmospheric escape in sub-Neptunes is thought to be driven by photoevaporation \citep{2013ApJ...775..105O} and/or core-powered mass loss \citep{2018MNRAS.476..759G}. At its orbital separation, the gravity of planet b would be sufficient to retain a H/He dominated envelope if stellar irradiation is the main driver \citep{2018ApJ...853..163J}. Should the large mass be confirmed with radial velocities, it could favour core-powered mass loss as the main driver atmospheric escape in this scenario, as the heat source is the planet's own residual formation energy \citep{2020MNRAS.493..792G}.

Given the two outer planets' relatively large orbital radii ($>50$ days), it is unlikely that they suffered major mass loss \citep{2020MNRAS.493..792G}. Preliminary models of the contraction of the planetary envelope based on the approach used by \citet{2006AAA...453L..21G} and \citet{2010AAA...516A..20V} and neglecting the core luminosity indicate that matching their mass, radius and age require a hydrogen-helium envelope which consists of 19\% of the planetary radius for planet c and 10\% for planet d. Such a relatively significant envelope mass fraction may result from a formation at large orbital distances \citep{2016ApJ...817...90L}. These values are to be considered with caution however because including core luminosity and tidal heating that may be caused by the resonances would yield smaller envelopes \citep{2019ApJ...886...72M}. 

Planets with vastly different densities within the same system, which is likely to be the case for TOI-282, provide invaluable laboratories to explore the various outcomes of planet formation with the same conditions.

\subsection{Prospects For Precise Mass Measurements With Radial Velocities}

\textcolor{black}{Stellar activity in HD~28109 poses a challenge on RV follow-up observations given the small RV semi-amplitudes of these planets. In our RV analysis, we find that the RV jitter terms in the HARPS and ESPRESSO measurements are on the order of $\rm 2.4\,m/s$ and $\rm 0.4\,m/s$ , respectively (Table~\ref{tab:glob_fit}).} Given the TTV masses of planets c and d, their expected RV semi-amplitudes are $\rm K_c \approx 1.14$~m/s and $\rm K_d \approx 0.71$~m/s. For planet b, the range is broad, spanning from the \cite{2017ApJ...834...17C} estimate to the mass of a pure iron planet, $\rm K \approx 0.91 - 2.4$~m/s. Given the length of the period of the outermost planet, sampling the full orbit will be a considerable undertaking.  

\textcolor{black}{With the goal to further characterize the system, we are currently conducting a radial velocity survey with ESPRESSO;} once concluded, the results will be presented in a follow-up paper. These high-precision radial velocity measurements will help to disentangle the eccentricity of planet b, as well as shed light on the possible presence of additional non-transiting planets in the system. In addition, these measurements will finally put precise constraints on the masses of these planets, which will be crucial for atmospheric follow-up observations.



\subsection{Dynamical Exploration of Planet b's Parameters}
\label{sec:stable}

Our dynamical and photometric fits provided us with TTV mass estimates for the two outermost planets, along with strong evidence for circular orbits. However, the mass and eccentricity of planet b remain largely unconstrained, and a dynamical analysis might be able to restrict plausible parameters of planet b. The first 3-planet photometric fit in Section \ref{sec:glob_phot} found an eccentricity of $\sim$0.8, while the dynamical fits presented in Section \ref{sec:nesvorny} favour an eccentricity of $\sim$0.17. 

We use the Mean Exponential Growth factor of Nearby Orbits chaos index \citep[$Y(t)$, MEGNO;][]{2000AAAS..147..205C}  to assess the stability of the HD~28109 system with the full range of eccentricity and mass estimates for planet b. MEGNO is used to evaluate whether a body's trajectory will be stable following perturbations of its initial conditions. The time-averaged MEGNO $\langle Y(t) \rangle$ will tend to 2 for $t \rightarrow \infty$ if the motion of the body is quasi-periodic, while chaotic behaviour will cause $\langle Y(t) \rangle$ to tend to infinity instead.

We made a $50\times50$ mass-eccentricity grid for planet b with eccentricity values evenly spaced between $0-0.8$, and masses evenly spaced between $5-30~\rm M_{\oplus}$. We used the MEGNO implementation of the N-body integrator \textsc{Rebound} \citep{2012AAA...537A.128R} which in turn uses the Wisdom-Holman \textsc{WHFast} code \citep{2015MNRAS.452..376R}. We used steps of $1/50\rm th$ of the period of planet b (0.5 days) and integrated for 10\,000 orbits of the outermost planet ($\sim2\,300~\rm years$) to construct a two-dimensional MEGNO map. 
We find that the full mass range is non-chaotic but only for eccentricities smaller than $\sim$0.4; eccentricities up to $\sim$0.5 can be stable but only for the largest masses in the range we tested. This indicates that a wide range of mass values are possible for planet b, in line with the wide mass prior used for our dynamical and RV fits. 


\subsection{Prospects For Atmospheric Characterisation with {\it HST} and {\it JWST}}

To assess the suitability of HD~28109~b, c and d for atmospheric characterisation using transmission and emission spectroscopy, we calculated for each planet the Transmission and Emission Spectroscopy Metrics following \cite{2018PASP..130k4401K}. The TSMs for each planet are 20, 64, and 38 respectively, while the ESMs are 1.08, 1.43 and 0.52. Given the brightness of the host and the long periods of the planets leading to equilibrium temperatures below 900K, the secondary eclipses would have very low SNR, making characterisation with this method too challenging. However, their TSMs indicate that characterisation with transmission spectroscopy is possible in principle. 

\begin{figure}
    \centering
    \includegraphics[width = \columnwidth]{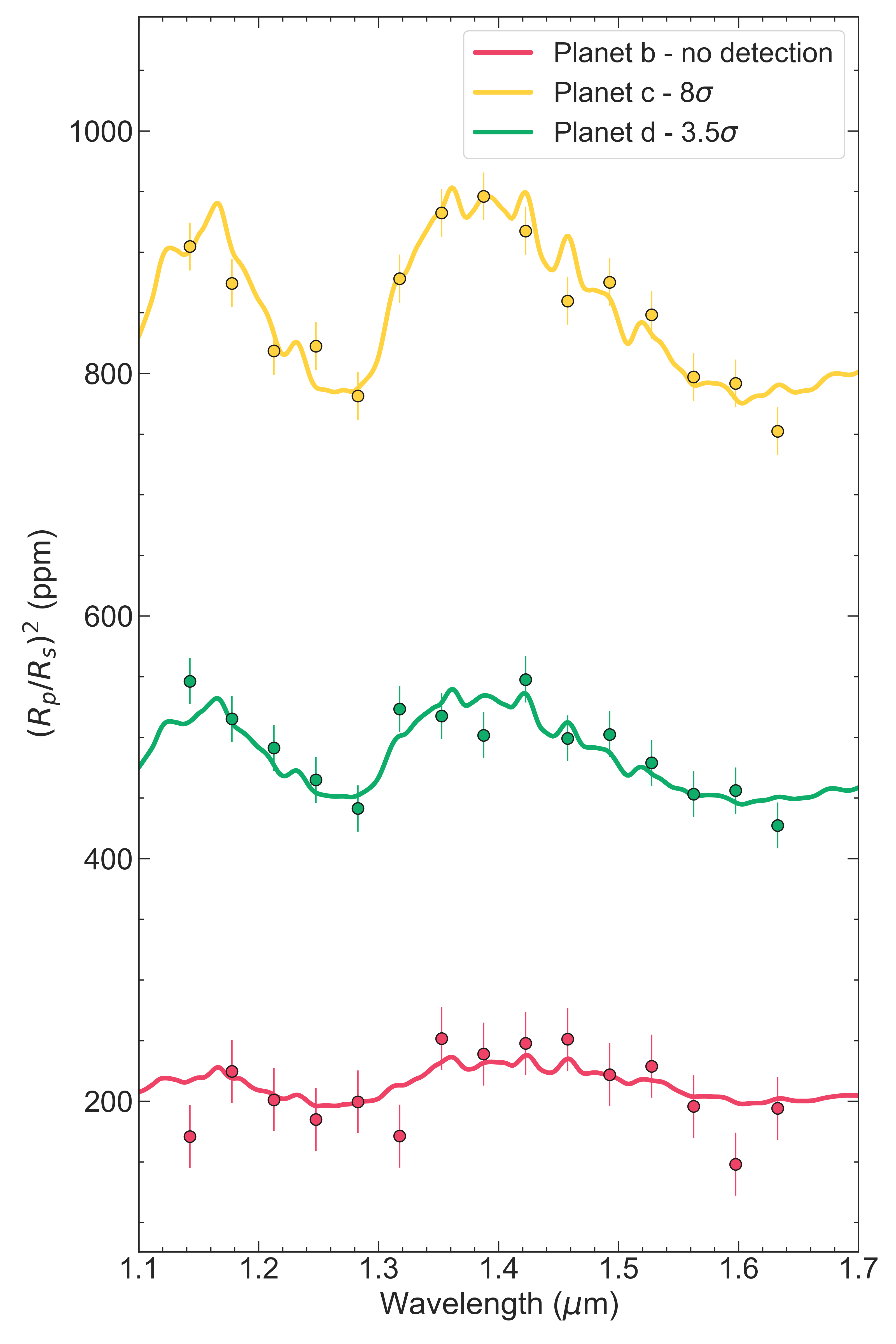}
    \caption{Model transmission spectra of HD~28109~b, c and d with simulated {\it HST} observations. Each point corresponds to a single transit depth with no offset; the curves represent the simulated transmission spectra. }
    \label{fig:hst}
\end{figure}

We produced model transmission spectra for all three planets using \textsc{Exo\_Transmit} \citep{2017PASP..129d4402K} assuming solar metallicity and C/O ratio for the host, and cloud-free H$_2$ dominated atmospheres for the planets. Using these spectra we then used \textsc{PandExo} \citep{2017PASP..129f4501B} to simulate Hubble Space Telescope ({\it HST} hereafter) observations of their atmospheres using transmission spectroscopy. Our simulations indicate that for the two outermost planets a single {\it HST} transit would be sufficient to carry out reconnaissance spectroscopy \citep[as in][]{2018NatAs...2..214D,2016Natur.537...69D}, to assess the presence of clouds in their atmospheres. In the event of a cloud-free atmosphere, we could detect the presence of water in transmission to $8\sigma$ and $3.5\sigma$ respectively. To reach a $3.5\sigma$ detection for planet b could require a large number of transits, but with its mass unconstrained for the time being we cannot make a fair estimate. As an example, should the mass be consistent with the \cite{2017ApJ...834...17C} estimate, we would require 6 transits, each spanning 8 orbits of {\it HST} to reach a $3.5\sigma$ detection. We present the model spectra along with simulated {\it HST} observations in Figure \ref{fig:hst}. 

We do also note that the simulations described above assume the TTV masses for planets c and d, and the mass estimated from mass-radius relations for planet b. Should any mass be significantly larger the scale height of the atmosphere will decrease, leading to lower significance detections. 

\section{Conclusions}
\label{sec:282conc}

In this work we have presented the discovery and confirmation of three sub-Neptune sized planets orbiting HD~28109. Given the brightness of the host, the planets are well-suited to precise radial velocity measurements, which will allow comparison of TTV masses with RV masses. This system is also the brightest \tess system discovered to date known to host TTV, making it an exquisite laboratory for planet formation theory. Additionally, we have shown that the outer two planets are well suited to atmospheric reconnaissance with current instrumentation, and they will be high-priority planets for follow-up with \jwst and \ariel.

This work has also demonstrated the capabilities of \astep as a observatory specialising in planets with long periods, very long transits, and large TTVs. The uninterrupted night enjoyed from Antarctica during two months of the year enables observations of systems that cannot be feasibly followed from anywhere else in the world. Continued selection of these candidates for follow-up will help the community fill out the most sparsely populated parameter spaces in the ever-growing field of exoplanets. 



\section*{Acknowledgements}

\textcolor{black}{The authors wish the thank the reviewer for their comments and feedback which helped to improve and clarify the manuscript.}
This work makes use of observations from the ASTEP telescope. ASTEP benefited from the support of the French and Italian polar agencies IPEV and PNRA in the framework of the Concordia station program, from OCA, INSU, and from Idex UCAJEDI (ANR-15-IDEX-01). The authors would also like the winterover staff of Concordia station; their work makes the continued operations of ASTEP possible. Some of the observations in the paper made use of the High-Resolution Imaging instrument Zorro obtained under Gemini LLP Proposal Number: GN/S-2021A-LP-105. Zorro was funded by the NASA Exoplanet Exploration Program and built at the NASA Ames Research Center by Steve B. Howell, Nic Scott, Elliott P. Horch, and Emmett Quigley. Zorro was mounted on the Gemini North (and/or South) telescope of the international Gemini Observatory, a program of NSF’s OIR Lab, which is managed by the Association of Universities for Research in Astronomy (AURA) under a cooperative agreement with the National Science Foundation. on behalf of the Gemini partnership: the National Science Foundation (United States), National Research Council (Canada), Agencia Nacional de Investigación y Desarrollo (Chile), Ministerio de Ciencia, Tecnología e Innovación (Argentina), Ministério da Ciência, Tecnologia, Inovações e Comunicações (Brazil), and Korea Astronomy and Space Science Institute (Republic of Korea).
\textcolor{black}{This work has made use of data from the European Space Agency (ESA) mission {\it Gaia} (\url{https://www.cosmos.esa.int/gaia}), processed by the {\it Gaia} Data Processing and Analysis Consortium (DPAC, \url{https://www.cosmos.esa.int/web/gaia/dpac/consortium}). Funding for the DPAC has been provided by national institutions, in particular the institutions participating in the {\it Gaia} Multilateral Agreement. 
This paper includes data collected by the \tess mission. Funding for the \tess mission is provided by the NASA's Science Mission Directorate.}
Based in part on observations collected at the European Organisation for Astronomical Research in the Southern Hemisphere under ESO programme(s) 0102.C-0503(A).
This work makes use of observations from the LCOGT network. Part of the LCOGT telescope time was granted by NOIRLab through the Mid-Scale Innovations Program (MSIP). MSIP is funded by NSF.
MNG acknowledges support from the European Space Agency (ESA) as an ESA Research Fellow.
Part of this work has been carried out within the framework of the NCCR PlanetS supported by the Swiss National Science Foundation.
This research received funding from the European Research Council (ERC) under the European Union's Horizon 2020 research and innovation programme (grant agreement n$^\circ$ 803193/BEBOP), and from the Science and Technology Facilities Council (STFC; grant n$^\circ$ ST/S00193X/1). DG gratefully acknowledges financial support from the Cassa di Risparmio di Torino (CRT) foundation under Grant No. 2018.2323 ``Gaseous or rocky? Unveiling the nature of small worlds''.
This research made use of {\sc Lightkurve}, a Python package for \Kepler and \tess data analysis \citep{2018ascl.soft12013L}, which in turn makes use of {\sc Astropy},\footnote{http://www.astropy.org} a community-developed core Python package for Astronomy \citep{2013AAA...558A..33A, 2018AJ....156..123A}, {\sc Astroquery} \citep{2019AJ....157...98G}, and {\sc TESSCut} \citep{2019ascl.soft05007B}. 

\section*{Data Availability}

\tess data products are available via the MAST portal at \url{https://mast.stsci.edu/portal/Mashup/Clients/Mast/Portal.html}. Follow-up photometry and high resolution imaging data for HD~28109 are available on ExoFOP at \url{https://exofop.ipac.caltech.edu/tess/target.php?id=29781292}. These data are freely accessible to ExoFOP members immediately and are publicly available following a one-year proprietary period. The radial velocity data used in this paper are available via the ESO Public Archive at \url{http://archive.eso.org/eso/eso_archive_main.html}.
 



\bibliographystyle{mnras}
\bibliography{TOI282} 





\appendix


\bsp	
\label{lastpage}
\end{document}